\documentclass[%
 reprint,
 amsmath,amssymb,
 longbibliography,
 aps,
 pre,
]{revtex4-2}

\usepackage{lipsum}
\usepackage{calc}
\usepackage{fancyhdr}

\usepackage[normalem]{ulem}
\usepackage{graphicx}
\usepackage{bm}
\usepackage{xcolor}
\usepackage{hyperref}
\usepackage{amsmath}
\usepackage{amssymb}
\usepackage[utf8]{inputenc}
\usepackage{ulem}
\usepackage{xr}
\usepackage{outlines}
\usepackage{subfigure}
\usepackage{bbm}
\usepackage{mathtools}
\usepackage{mathrsfs}
\usepackage{float}

\def\env{\bm{\xi}}

\def\Z{\mathbb{Z}}

\def\randMeasure{\nu}
\def\expEnv{\mathbb{E}^{\env}}

\newcommand{\expAnnealed}[1]{\mathbb{E}_{\nu}\left[ #1 \right]}

\def\varAnnealed{\mathrm{Var}_{\nu}}

\newcommand{\tiltedExpAnnealed}[1]{\tilde{\mathbf{E}}\left[ #1 \right]}
\newcommand{\tiltedExpAnnealedTwo}[1]{\tilde{\mathbf{E}}\left[ #1 \right]}
\def\extCoef{\lambda_\mathrm{ext}}
\newcommand{\dExt}[1]{D_{\mathrm{ext}}^{(#1)}}

\def\centeringTerm{c(N)}
\def\momentVar{\varAnnealed\left(\expEnv[Y^m]\right)}

\newcommand{\diffRandomWalk}[1]{\Delta\left( #1 \right)}
\newcommand{\absMeasure}[1]{\tilde{\mu}\left( #1 \right)}

\newcommand{\meanOmega}[1]{\bar{\xi}(#1)}
\newcommand{\invMeasure}[1]{\mu({#1})}

\newcommand{\var}[1]{\mathrm{Var}\left( {#1} \right)}

\def\envMax{\mathrm{Env}_t^N}
\def\max{\mathrm{Max}_t^N}
\def\samMax{\mathrm{Sam}_t^N}

\def\envFPT{\mathrm{Env}_L^N}
\def\tauL{\tau_L}
\def\min{\mathrm{Min}_L^N}
\def\samFPT{\mathrm{Sam}_L^N}

\begin{document}

\title{Super-Universal Behavior of Outliers Diffusing in a Space-Time Random Environment}
\author{Jacob Hass}
\affiliation{Department of Physics and Materials Science Institute, University of Oregon, Eugene, Oregon 97403, USA.}
\date{\today}
\begin{abstract}
I characterize the extreme location and extreme first passage time of a system of $N$ particles independently diffusing in a space-time random environment. I show these extreme statistics are governed by the Kardar-Parisi-Zhang (KPZ) equation and derive their mean and variance. I find the scalings of the statistics depend on the moments of the environment. Each scaling regime forms a universality class which is controlled by the lowest order moment which exhibits random fluctuations. When the first moment is random, the environment plays the role of a random velocity field. When the first moment is fixed but the second moment is random, the environment manifests as fluctuations in the diffusion coefficient. As each higher moment is fixed, the next moment determines the scaling behavior. Since each scaling regime forms a universality class, this model for diffusion forms a super-universality class. I confirm my theoretical predictions using numerics for a wide class of underlying environments. 

\end{abstract}

\maketitle

\emph{Introduction.} The classical theory of diffusion reduces the stochastic and chaotic motion of particles due to a complex environment to a single parameter -- the diffusion coefficient, $D$. \cite{einsteinTheoretischeBemerkungenUber1907a, einsteinUberMolekularkinetischenTheorie1905a, einsteinZurTheorieBrownschen1906a, sutherlandViscosityGasesMolecular1893a, vonsmoluchowskiZurKinetischenTheorie1906a}. This simplification relies on two key assumptions: particles evolve independently and the mean and variance of their jump distribution are well defined. When the second assumption is violated, specifically when the variance is not well defined, this leads to phenomenon described by anomalous diffusion and L\'evy flights \cite{chechkinIntroductionTheoryLevy2008, sokolovDiffusionAnomalousDiffusion2005, metzlerRandomWalksGuide2000, bouchaudAnomalousDiffusionDisordered1990a}. I study \emph{Random Walks in a Random Environment} (RWRE) models which violates the first assumption, independence, by introducing a space-time random environment that induces correlations in the movement of nearby particles. Here I show the environment effects the fluctuations of particles at the edge of a system of $N$ diffusing particles. I show that the fluctuations in these extreme particles are characterized by the Kardar-Parisi-Zhang (KPZ) equation \cite{kardarDynamicScalingGrowing1986a, corwinKardarParisiZhang2012a} with universal scalings. In doing so, I derive the prefactors through which the environment controls these statistics. Even if the leading moments of the environment are fixed the higher order moments will control the fluctuations at the edge of the system. 

\begin{figure}[H]
\centering
\includegraphics[width=0.85\columnwidth]{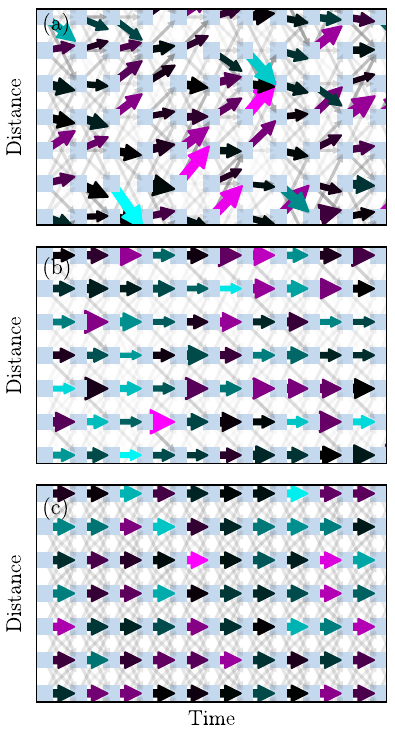}
\vspace{-0.3cm}
\caption{Several examples of random environments. Blue boxes are lattice sites and the opacity of the gray arrows shows the probability of jumping to a site. Large, solid arrows show the average drift at each site. The size of the arrow scales with the variance of the jump distribution and the color denotes the skew (cyan skewed down and pink skewed up). (a) Environment where the drift, variance and skewness are random. (b) Environment with no drift, but the variance and skewness are random. (c) Environment with no drift, a constant variance but a random skewness.}
\label{fig:Environments}
\end{figure}

I study RWRE models for diffusion wherein particles evolve on the integer lattice in a space-time random environment as shown in Fig. \ref{fig:Environments}. I derive the mean and variance of the first time a particle passes a barrier at position $L$, the \emph{extreme first passage time}, and the maximum position at time $t$, the \emph{extreme location}, for a system of $N$ diffusing particles. Previous work has shown that when the first moment of the environment is random, the distribution of these extreme statistics form a universality class which is governed by the KPZ universality class \cite{hassAnomalousFluctuationsExtremes2023, hassFirstpassageTimeManyparticle2024, hassExtremeDiffusionMeasures2024a, hassUniversalKPZFluctuations2025, ledoussalDiffusionTimeDependentRandom2017c, barraquandModerateDeviationsDiffusion2020c, brockingtonEdgeCloudBrownian2022a}. Here, I generalize these results to show there are an infinite number of universality classes where the lowest random moment of the environment determines which universality class a RWRE model is governed by. 


\emph{RWRE Models.} 
I define the class of RWRE models that I study. I begin by defining the space-time random environment as follows. Let $\randMeasure$ define a probability distribution on the space of probability distributions on the integers, $\Z$. I define $\xi$ as a probability distribution on $\Z$ which is sampled according to $\randMeasure$. Then $\xi(i)$ denotes the probability mass at $i \in \Z$. Each choice of $\randMeasure$ defines a new RWRE model. 

Given a  distribution $\randMeasure$, I define the environment as $\env = \{\xi_{x,t}: x \in \Z, t \in \Z_{>0}\}$ where each $\xi_{x,t}$ is a probability distribution on $\Z$ independently sampled from $\randMeasure$ for each site $(x, t)$. Three examples of an environment, $\env$, are given in Fig. \ref{fig:Environments}, where $\xi_{x,t}$ is given by the light grey arrows at each $(x,t)$. Physically, a single instatiation of an experiment of particles diffusing yields a different environment, $\env$. 


I give an example of $\randMeasure$ where the first moment of the jump distribution, $\sum_{i \in \Z} \xi_{x,t}(i) i$, is random and another example where the first moment is fixed. Fixing higher moments becomes more nuanced, but I have given several examples in the ``End Matter``. To achieve an environment where the first moment of the jump distribution is random, I use the \emph{Dirichlet} distribution which is sampled as follows. Begin by generating a list of Gamma distributed random variables, $(X_{-k}, \ldots, X_{k})$, with shapes $\vec{\alpha} = (\alpha_{-k}, \ldots, \alpha_{k})$ for $\alpha_i > 0$ and rate $\beta=1$ where $k\in \Z_{>0}$ determines the width of the distribution. A sample of the Dirichlet distribution is this list of Gamma distributed random variables normalized to sum to $1$ by dividing by their sum, $Z = \sum_{i=-k}^k X_i$. Then set $\xi(i) = X_i / Z$ for $i \in [-k, k]$ and $\xi=0$ otherwise. To achieve an environment where the first moment is fixed, but the second moment is random, I study the \emph{symmetric Dirichlet} distribution. A sample of this distribution is generated by sampling the Dirichlet distribution, $\frac{1}{Z}(X_{-k}, \ldots, X_{k})$, for a given $\vec{\alpha}$ and then setting $\xi(i) = (X_{i} + X_{-i})/(2Z)$ for $i\in[-k, k]$ and $\xi=0$ otherwise. Thus, the first moment is always zero, i.e. $\sum_{i \in \Z} \xi_{x,t}(i) i =0$, but the second moment is random. 

I define $\mathbb{E}_\randMeasure[\bullet]$ and $\mathrm{Var}_\randMeasure(\bullet)$ as the expectation and variance over the probability distribution $\randMeasure$. This expectation and variance is obtained by taking the average over many different environments, $\env$, sampled via $\randMeasure$. I denote $\mathbb{E}_\randMeasure[\xi]$ as the average probability distribution where I have dropped the subscript $x,t$ because all $\xi_{x,t}$ are sampled independently from $\randMeasure$. Thus, the average of each $\xi_{x,t}$ is the same. I will only consider distributions $\randMeasure$ which produce an average of no drift, i.e. $\sum_{i \in \Z} \mathbb{E}_\randMeasure[\xi(i)] i = 0$; although, my results likely hold for models with drift after shifting to a moving reference frame. 

I define $\mathbb{P}^{\env}$ as the probability measure of $N$ independent random walks, $R^1, \ldots, R^N$, evolving in the same environment, $\env$. I let $\mathbb{E}^{\env}[\bullet]$ and $\mathrm{Var}^{\env}(\bullet)$ denote the mean and variance associated with the measure $\mathbb{P}^{\env}$. This mean and variance corresponds to averaging over many samples of a random variable in the \emph{same} environment. I denote $R^i(t)\in \Z$ as the position of the $i^{\mathrm{th}}$ walker at time $t$ and set $R^i(0) = 0$ for all $i\in[1, N]$ so all walkers start at the origin. The transition probabilities of each walker in $\env$ is given by 
\begin{equation*}
\mathbb{P}^{\env}(R(t+1) = x+i \mid R(t) = x) = \xi_{x,t}(i)
\end{equation*}
where I have dropped the super-script $i$ here and below since all walkers are independent and identically distributed (i.i.d). The walkers evolve independently in the same environment, but walks at the same site $(x, t)$ have correlated motion since they sample the same jump distribution, $\xi_{x,t}$. 

Given a single random walk, I denote the probability of the walk being at site $x$ at time $t$ as $p^{\env}(x,t) \coloneqq \mathbb{P}^{\env}(R(t) = x)$. By independence, all random walks have the same distribution $p^{\env}(x,t)$. This distribution obeys the recurrence relation 
\begin{equation*}
p^{\env}(x, t+1) = \sum_{j \in \Z} p^{\env}(x-j, t) \xi_{t, x-j}(j)
\end{equation*}
with initial condition $p^{\env}(x, 0) = \delta_{x,0}$ where $\delta_{x,0}$ is the Kronecker-delta. 

I conclude by defining the mean and variance of a random variable averaged over many samples from many different environments which I denote as $\mathbf{E}[\bullet]$ and $\mathbf{Var}(\bullet)$. These are the mean and variance which would be experimentally measurable, as these statistics correspond to averaging over many runs of an experiment where the random, underlying environment is unknown. I relate $\mathbf{E}[\bullet]$ and $\mathbf{Var}(\bullet)$ to my previous definitions using the total law of expectation and variance such that $\mathbf{E}[\bullet] = \mathbb{E}_\randMeasure[\mathbb{E}^{\env}[\bullet]]$ and $\mathbf{Var}(\bullet) = \mathbb{E}_\randMeasure[\mathrm{Var}^{\env}(\bullet)] + \mathrm{Var}_\randMeasure(\mathbb{E}^{\env}[\bullet]).$

\emph{Extreme Value Statistics.} 
For a system of $N$ particles, I study the extreme location and the extreme first passage time. I define the extreme location at time $t$ as 
\begin{equation*}
	\max \coloneqq \mathrm{max}(R^1(t), \ldots, R^N(t)).
\end{equation*}
Since the random walks are independent, the distribution of $\max$ in $\env$ is given by
\begin{equation}\label{eq:maxDist}
\mathbb{P}^{\env}(\max \leq x) = (1- \mathbb{P}^{\env}(R(t) \geq x))^N.
\end{equation}

I define $\tau_L^{i}$ as the first time the $i^{\mathrm{th}}$ particle passes position $L \in \Z_{>0}$. Then the extreme first passage time of $N$ particles at a distance $L$ is 
\begin{equation*}
\min \coloneqq \mathrm{min}(\tau_L^1, \ldots, \tau_L^N).
\end{equation*} 
Since the location of the random walks are i.i.d., all $\tau_L^i$ are also i.i.d. Thus, the distribution of the extreme first passage time is 
\begin{equation}\label{eq:minDist}
	\mathbb{P}^{\env}(\min \geq t) = (1- \mathbb{P}^{\env}(\tauL \leq t))^N.
\end{equation}

I derive $\mathbf{E}[\bullet]$ and $\mathbf{Var}(\bullet)$ for $\max$ and $\min$; that is, the mean and variance calculated over many experiments in different environments. I break each extreme value statistic into two sources of randomness: the mean of the statistic in a given environment and fluctuations about that mean. 

I begin by discussing the extreme location. I approximate the mean of the extreme location in a given environment, $\mathbb{E}^{\env}[\max]$, with $\envMax$. I define $\envMax$ as 
\begin{equation*}
\envMax \coloneqq \mathrm{max}\left(x: \mathbb{P}^{\env}(R(t) \geq x) \geq \frac{1}{N}\right).
\end{equation*}
I argue $\envMax$ is a natural approximation for $\mathbb{E}^{\env}[\max]$ because by combining Eq. \ref{eq:maxDist} with my definition for $\envMax$, $\mathbb{P}^{\env}(\max \leq \envMax) \approx e^{-1}$ as $N \rightarrow \infty$. Therefore, $\envMax$ is approximately the median of $\max$ in a given environment which can further be approximated as $\mathbb{E}^{\env}[\max]$. Notice $\envMax$ only captures environmental fluctuations since it only depends on $\env$. For classical diffusion, $\envMax$ is constant since the environment is constant. I define the fluctuations about $\envMax$ as $\samMax \coloneqq \max - \envMax$. Then $\samMax$ represents the effects of sampling $\max$ in a given environment. Physically, the distribution of $\samMax$ can be constructed by measuring $\max$ many times in the $\emph{same}$ environment and then subtracting the measured mean, i.e. $\mathbb{E}^{\env}[\max]$. 

Similarly, for the extreme first passage time, I approximate $\mathbb{E}^{\env}[\min]$ via 
\begin{equation*}
\envFPT \coloneqq \mathrm{min}\left(t: \mathbb{P}^{\env}(\tau_L \leq t) \geq \frac{1}{N} \right) 
\end{equation*}
and fluctuations about $\envFPT$ as $\samFPT \coloneqq \min - \envFPT$. 

\emph{Theoretical Methods.}
I generalize the methods presented in \cite{hassExtremeDiffusionMeasures2024a} to characterize the tail probability, $\mathbb{P}^{\env}(R(t) \geq x)$, in the limit that $t \rightarrow \infty$. I use the replica method to derive the first two moments of the tail probability and show the moments converge to those of the multiplicative stochastic heat equation (SHE),
\begin{equation*}
\partial_T Z(X,T) = \partial_X^2 Z(X, T) + \sqrt{2D_0} Z(X, T) \eta  
\end{equation*}
where $Z(X,T)$ is the solution to the SHE, $D_0 \in \mathbb{R}_{>0}$ is the noise strength and $\eta$ is space time white noise with the properties $\mathbb{E}[\eta(X, T)] = 0$, $\mathbb{E}[\eta(X, T) \eta(X', T')] = \delta(X-X', T-T')$ where $\delta(x,t)$ is the Dirac delta function. Here, $\mathbb{E}[\bullet]$ represents and average over the noise, $\eta$. The KPZ equation is related to the SHE via a logarithm , $Z(X,T) = e^{h(X, T)}$, where $h(X, T)$ solves the KPZ equation
\begin{equation*}
\partial_T h(X,T) = \partial_X^2 h(X, T) + \left(\partial_X h(X, T)\right)^2 + \sqrt{2D_0} \eta.
\end{equation*}
The crux of my derivation is identifying the noise strength of the SHE in terms of the statistics of the environment. See \cite{SeeSupplementalMateriala} for the derivation of the first two moments.

The scaling regime where these KPZ fluctuations appear scales as $R(t) \propto t^{\frac{4m-1}{4m}}$ where $m \in \Z_{>0}$ and $m-1$ is the number of deterministic moments of the jump distribution \cite{t78Regime}. When $m=1$, there are no moments fixed in the environment, as shown in Fig. \ref{fig:Environments}a. This corresponds to an environment which acts as a space-time random velocity field creating correlations in the velocity of nearby particles. When $m=2$, the first moment of the environment is always fixed such that $\nu$ only generates distributions where $\sum_{i\in\Z} \xi_{x,t}(i) i = 0$, as shown in Fig. \ref{fig:Environments}b. Since I only consider net drift free environments, the first moment must always be 0. This corresponds to an environment where there are no correlations in the velocity of nearby particles, but the diffusion coefficient is a space-time random value. When $m=3$, the first moment and second moments are deterministic such that $\nu$ only generates distributions where $\sum_{i\in\Z} \xi_{x,t}(i) i = 0$ and $\sum_{i \in \Z} \xi_{x,t}(i) i^2 = 2 D$ for $D \in \mathbb{R}_{>0}$, as shown in Fig. \ref{fig:Environments}c. 

%

\begin{figure*}[ht]
\hspace{-3cm}
\begin{minipage}{.4\textwidth}
\includegraphics[scale=0.6]{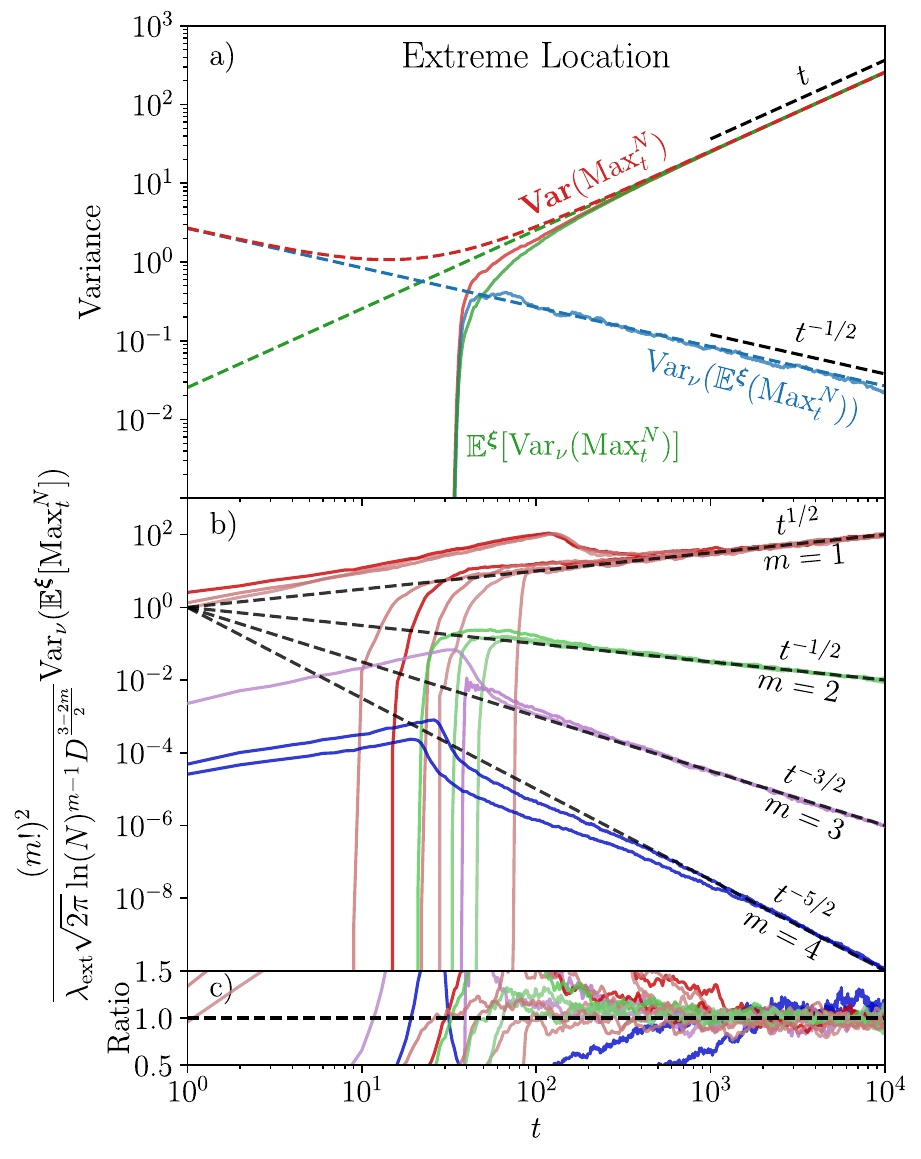}
\end{minipage}\qquad\qquad\qquad
\begin{minipage}{.4\textwidth}
\includegraphics[scale=0.6]{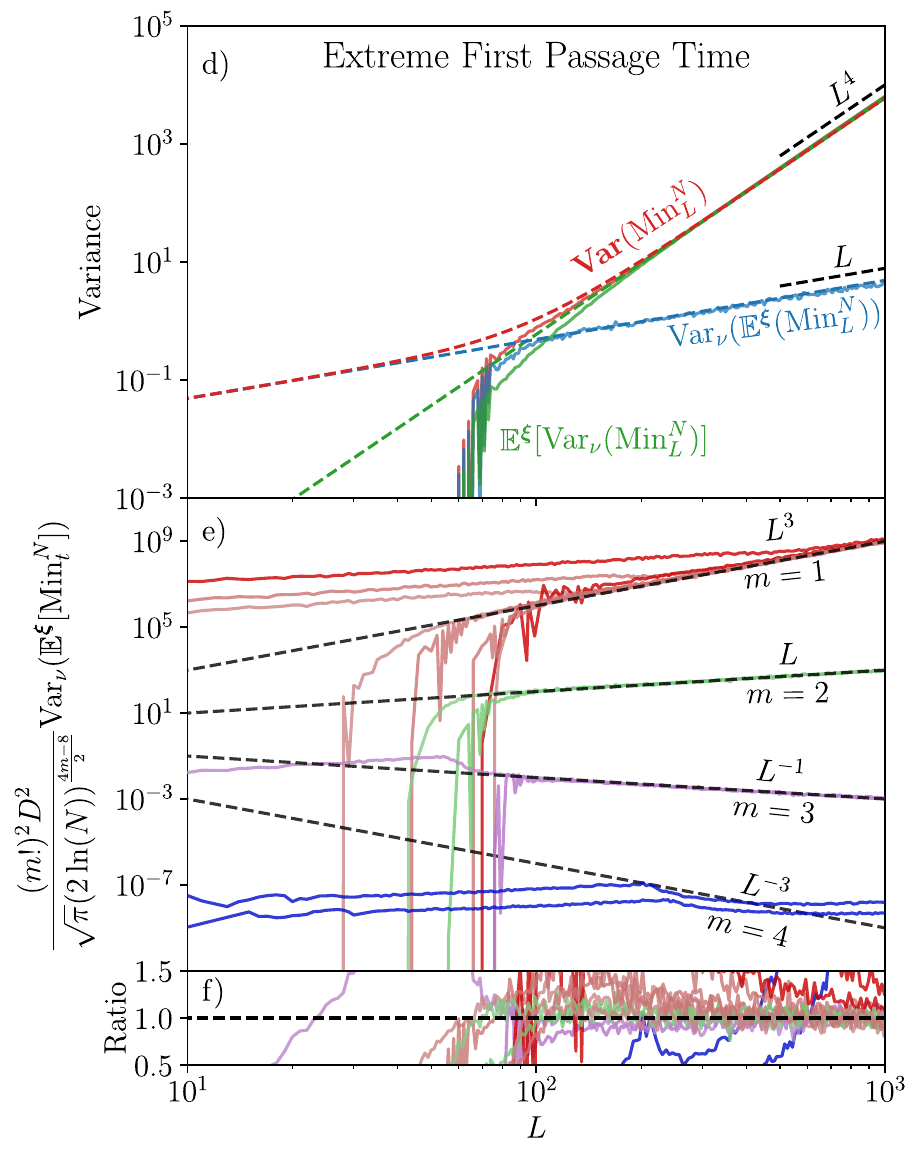}
\end{minipage}
\caption{Plot of the different variances for the extreme location (a) and the extreme first passage time (d), for the symmetric Dirichlet distribution and systems of size $N=10^{28}$. Dashed lines represent my theoretical predictions and solid lines are numerically measured values. Collapse of the environmental fluctuations of the extreme location (b) and extreme first passage time (e) for $m=1$ (red), $m=2$ (green), $m=3$ (purple) and $m=4$ (blue). I plot a number of different distributions listed in the ``End Matter". The saturation of each curve is scaled by the diffusion coefficient, $D$. The ratio of measured $\envMax$ (c) and $\envFPT$ (f) to their theoretical predictions in Eq. \ref{eq:envMax} and \ref{eq:envFPT}, respectively. All curves are computed over 500 independent environments.}
  \label{fig:FPTMax}
\end{figure*}

\emph{Theoretical Results.}
I define the diffusion coefficient for a RWRE model as half the variance in the average environment,
\begin{equation}\label{eq:diffusionCoefficient}
	D \coloneqq \frac{1}{2} \sum_{i \in \Z} \mathbb{E}_{\randMeasure}[\xi(i)] i^2
\end{equation}
where the mean in the average environment is 0 since I consider net drift free environments. Since the tail probability depends on the number of moments that are fixed, $m$, the extreme value statistics will as well. I derive the following results for the extreme location and extreme first passage time in the limit that $t^{2m-1} \gg \extCoef^2 \ln(N)^{2m}$ and $L^{4m-2} \gg \extCoef^2 \ln(N)^{4m-1}$, respectively, where $\extCoef$ is defined in Eq. \ref{eq:extCoef} and $N \gg 1$. 

I find the mean of $\max$ and $\min$ are given by 
\begin{equation*}
\mathbf{E}[\max] \approx \sqrt{4 D t \ln(N)} \quad \mathrm{and} \quad \mathbf{E}[\min] \approx \frac{L^2}{4D\ln(N)}
\end{equation*}
which match the classical predictions \cite{madridCompetitionSlowFast2020a, lawleyDistributionExtremeFirst2020a, linnExtremeHittingProbabilities2022a, lawleyUniversalFormulaExtreme2020a, basnayakeAsymptoticFormulasExtreme2019a, schussRedundancyPrincipleRole2019a, rednerRedundancyExtremeStatistics2019a}. Notice that the means only depend on the diffusion coefficient of the model and not any other statistics of the environment.  

Although the means give the same results as classical predictions, the variances display more interesting behavior. I find the variance of $\samMax$ and $\samFPT$ are Gumbel distributed with variances
\begin{align}
\mathbf{Var}(\samMax) &\approx \frac{\pi^2 D t}{6 \ln(N)} \\ \mathbf{Var}(\samFPT) &\approx \frac{\pi^2 L^4}{96 D^2 \ln(N)^4}.
\end{align}
These results match the variance of the extreme location and first passage time predicted classically \cite{madridCompetitionSlowFast2020a, lawleyDistributionExtremeFirst2020a, linnExtremeHittingProbabilities2022a, lawleyUniversalFormulaExtreme2020a, basnayakeAsymptoticFormulasExtreme2019a, schussRedundancyPrincipleRole2019a, rednerRedundancyExtremeStatistics2019a} and thus only depend on the diffusion coefficient and no other statistics of the environment. This makes sense because in classical diffusion, the environment is homogeneous and thus the variance only depends on fluctuations due to sampling random walks.

I find the variance of $\envMax$ and $\envFPT$ are given by 
\begin{align}
\varAnnealed(\envMax) &\approx \frac{\sqrt{2 \pi}}{(m!)^2} \extCoef \ln(N)^{m-1} (Dt)^{\frac{3-2m}{2}} \label{eq:envMax} \\
\varAnnealed(\envFPT) &\approx \frac{\extCoef \sqrt{\pi} 2^{\frac{4m-9}{2}}}{(m!)^2 D^2} \ln(N)^{\frac{4m-9}{2}} L^{5-2m} \label{eq:envFPT}
\end{align}
where $m-1$ is the number of deterministic moments and 
\begin{equation}\label{eq:extCoef}
\extCoef = \frac{\momentVar}{2\mathbb{E}_\randMeasure[\mathrm{Var}^{\env}(Y)]}
\end{equation}
where $Y \sim \xi_{x,t}$ is distributed according to a single step of a random walk such that 
\begin{align}
\momentVar &= \expAnnealed{\left( \sum_{i \in \Z} \xi_{x,t}(i) i^m \right)^2} \nonumber \\ & \quad \quad \quad \quad \quad - \left( \sum_{i \in \Z} \mathbb{E}_\randMeasure[\xi(i)] i^m \right)^2, \\
\mathbb{E}_\randMeasure[\mathrm{Var}^{\env}(Y)] &= 2D - \mathrm{Var}_\randMeasure(\mathbb{E}^{\env}[Y]).
\end{align}
Since $\envMax$ and $\envFPT$ are constant in a given environment, I calculate the mean and variance with respect to $\randMeasure$. 

Notice that for $m>1$, the environmental variance for the extreme location, $\varAnnealed(\envMax)$, decreases in $t$. Similarly, for $m>2$, the environmental variance for the extreme first passage time, $\varAnnealed(\envFPT)$, decreases in $L$. Thus, any RWRE model with $m>2$ will yield the same measurements, as $t\rightarrow \infty$ for the extreme location and $L\rightarrow\infty$ for the extreme first passage time, as the classical predictions since the environmental variance asymptotes to 0 but the sampling variance persists. Luckily, the $m=1$ and $m=2$ case have the most physical interpretations as they correspond to an environment which represents an underlying velocity field and random diffusion coefficient, respectively. 

For $m=1$, the form of $\extCoef$ in Eq. \ref{eq:extCoef} only contains one unknown quantity, $\frac{1}{2}\mathbb{E}_\randMeasure\left[\left( \sum_{i \in \Z} \xi_{x,t} i \right)^2\right]$, which was termed the \emph{extreme diffusion coefficient}, $D_\mathrm{ext}$, in \cite{hassExtremeDiffusionMeasures2024a}. For $m>1$, I define an analogous extreme diffusion coefficient as $\dExt{m} \coloneqq \frac{1}{2}\momentVar$. Thus, the extreme diffusion coefficient is simply the variance of the $m^{\mathrm{th}}$ moment of the jump distribution over all environments. For $m=1$, $\dExt{1}$ is the variance in the drift and for $m=2$, $\dExt{2}$ is the variance in the diffusion coefficient. 

I derive Eq. \ref{eq:extCoef} under the condition 
\begin{equation}\label{eq:condition}
\mathbb{E}_\randMeasure[\xi_{x,t}(i) \xi_{x,t}(j)] = c \mathbb{E}_\randMeasure[\xi_{x,t}(i)] \mathbb{E}_{\randMeasure}[\xi_{x,t}(j)] 
\end{equation}
for all $i \neq j$ where $c \in (0, 1)$. This condition is satisfied for every distribution I consider for $m=1$. Although this is not the case for the distributions that I consider for $m>1$, Eq. \ref{eq:condition} is approximately satisfied in the weak noise limit where $\xi_{x,t}(i) = \mathbb{E}_\randMeasure[\xi_{x,t}(i)] + \sigma_{x,t}(i)$ and $\sigma_{x,t}(i) \ll \mathbb{E}_\randMeasure[\xi_{x,t}(i)]$ for all $i \in \Z$.  

In finding the first two moments of $\envMax$ and $\samMax$, I find they are independent as $t\rightarrow\infty$. Similarly, $\envFPT$ and $\samFPT$ are independent as $L\rightarrow\infty$. By this independence, I find
\begin{align}
\mathbf{Var}(\max) &\approx \varAnnealed(\envMax) + \mathbf{Var}(\samMax) \label{eq:maxAddition} \\ 
\mathbf{Var}(\min) &\approx \varAnnealed(\envFPT) + \mathbf{Var}(\samFPT). \label{eq:fptAddition}
\end{align}

\emph{Numerical Results.}
Figs. 2a,d) show the collapse of each variance to their numerically measured counterpart in the limit that $t \rightarrow \infty$ and $L\rightarrow \infty$. I find $\varAnnealed(\mathbb{E}^{\env}[\max])$ is well approximated by $\varAnnealed(\envMax)$ and $\expAnnealed{\varAnnealed(\max)}$ is well approximated by $\varAnnealed(\samMax)$. The same is true for the analogous variances of the extreme first passage time. This collapse confirms the independence of the environmental and sampling variance and thus the addition laws in Eq. \ref{eq:maxAddition} and \ref{eq:fptAddition}. Figs. b,e) show the numerically measured environmental variance asymptotes to my theoretical predictions for a wide range of distributions and $m$. For $m\geq 2$, Eq. \ref{eq:condition} is not satisfied for any of the show distributions. However, I still use the simplified form of $\extCoef$ in Eq. \ref{eq:extCoef} under the assumption that these distributions are in the weak noise limit. This appears to be a reasonable approximation because my predictions match the numerics quite well. For the extreme first passage time and $m>3$, my theoretical predictions break down. This is because my predictions for the extreme first passage time are derived by translating the KPZ fluctuations at $R(t) \propto t^{\frac{4m-1}{4m}}$ to the bulk regime where $R(t) \propto \mathcal{O}(1)$. In the bulk regime, \cite{drillickRandomWalksSpacetime2025} predicts that there are Gaussian fluctuations with scaling behavior that is not captured by the short-time Gaussian statistics of the KPZ equation. 

\emph{Conclusion.}
I have shown the extreme value statistics of RWRE models for diffusion form a super-universality class which is governed by the KPZ equation. I find the variance of the extreme statistics display anomalous power laws given in Eq. \ref{eq:envMax} and \ref{eq:envFPT} not captured by classical diffusion. The prefactor of the statistics depends on a single unknown quantity , $\extCoef$, in Eq. \ref{eq:extCoef} which is defined in terms of the variance of the lowest random moment of the environment. Thus, a measurement of the extreme location or extreme first passage time could be used to measure microscopic fluctuations of the environment which are ignored in classical diffusion. The diffusive environment I study is quite general so I expect this to apply to a wide number of physical systems. This includes particles diffusing in a space-time random velocity field and systems with a space-time random diffusion coefficient.

\emph{Acknowledgments.} This work was funded by the W.M. Keck Foundation Science and Engineering grant on "Extreme Diffusion". This work benefited from access to the University of Oregon high performance computing cluster, Talapas. I would also like to thank Eric Corwin, Ivan Corwin, Hindy Drillick and Shalin Parekh for their helpful discussion.

\bibliographystyle{unsrt}
\bibliography{main}

\clearpage
\onecolumngrid
{\centering {\huge \textbf{End Matter} \par}}
\setcounter{section}{0}
\bigskip
\twocolumngrid 

\section{Distributions for $\randMeasure$ in Figure \ref{fig:FPTMax}}

Here I list the various distributions for $\randMeasure$ I used for the numerics in Fig. \ref{fig:FPTMax}. I begin by defining the distributions I study where the first moment is random, i.e. $m=1$. I then move onto $m=2,3$ and $4$ where I fix more moments of the jump distribution.

For $m=1$, I use the same distributions we studied in \cite{hassExtremeDiffusionMeasures2024a}. This includes the Dirichlet distribution for $\vec{\alpha}= (12, 1, 12)$ and $\vec{\alpha}=(2 , 1, 1/4, 4, 1/2)$. I also studied the \emph{flat Dirichlet distribution} which is the Dirichlet distribution with all $\alpha_i = 1$ on the interval $[-k, k]$. I used the flat Dirichlet distribution with $k=3, 5$ and $11$. 

I also studied the \emph{random delta} distribution defined as follows. The random delta distribution is controlled by a choice of $k \in \Z_{>0}$. Then the random delta distribution is sampled by drawing two numbers $X_1$ and $X_2$ uniformly and without replacement from $\{-k, \ldots k\}$. I then set $\xi(X_1) = \xi(X_2) = 1/2$ and $\xi(i) = 0$ for all $i \neq X_1, X_2$. I used the random delta distribution with parameter $k = 3, 5$ and $11$. 

For $m=2$, I use the symmetric Dirichlet distribution with $\vec{\alpha} = (1, 1, 1)$ and $\vec{\alpha} = (1, 1, 1, 1, 1)$.


For $m=3$, I study two distributions where the first moment is always zero, the second moment is constant, and the third moment is random. The first way I do this is to first construct a distribution which has a mean of zero and a finite second and third moment. Then by flipping the distribution, i.e. swapping $\xi_{x,t}(i)$ and $\xi_{x,t}(-i)$, randomly the mean and second moment stay the same, but the third moment changes sign. In this spirit, I define the \emph{randomly flipped Beta Binomial} distribution as follows. I define the vector
\begin{equation*}
\vec{X} = \left(\frac{1}{14}, \frac{3}{14}, \frac{5}{14}, \frac{5}{14}, 0 \right).
\end{equation*}
I then flip a coin, and if heads let $\xi_{x,t}(i) = \vec{X}(i)$ for $i \in [-2, 2]$ and tails let $\xi_{x,t}(i) = \vec{X}(-i)$. I set $\xi_{x,t}(i) = 0$ otherwise. This ensures $\xi_{x,t}$ has the properties $\sum_{i \in \Z} \xi_{x,t}(i) i = 0$, $\sum_{i \in \Z} \xi_{x,t}(i) i^2 = 6/7$ and $\sum_{i \in \Z} \xi_{x,t}(i) i^3 = \pm 3/7$ depending on if coin lands heads or tails. I call this the randomly flipped Beta Binomial distribution because $\vec{X}$, ignoring the last term that is $0$, is the probability mass function of the Beta-binomial distribution with number of trials $n=3$ and parameters $\alpha=3$ and $\beta=4$. 

Although the randomly flipped Beta Binomial distribution has a third moment which is random, I construct a more general distribution which has a random third moment. I first constrain the distribution $\xi_{x, t}$, to be nonzero on the interval $[-2,2]$, and  zero otherwise. I have three constraints on the distribution: $\sum_{i \in \Z} \xi_{x,t}(i) = 1$, $\sum_{i \in \Z} \xi_{x,t}(i) i = 0$ and $\sum_{i \in \Z} \xi_{x,t}(i) i^2 = 2D$ for $D > 0$. Since $\xi_{x,t}$ has fine non-zero values and I have three constraints, there are two degrees of freedom in the distribution. One way to interpret this, is that there are two random values in the distribution $\xi_{x,t}$ and the other elements are functions of these random values. For example, I define the \emph{random two DOF} distribution as 
\begin{equation*}
\xi_{x,t} = \left(\frac{1}{6}(2 - 3 a - b), a, b, \frac{1}{3}(2-3a-4b), \frac{a+b}{2} \right)
\end{equation*}
where $a$ and $b$ are uniform random variables on the interval $\left[\frac{2}{15}, \frac{4}{15} \right]$. This ensures that $\sum_{i \in \Z} \xi_{x,t}(i) i = 0$, $\sum_{i \in \Z} \xi_{x,t}(i) i^2 = 2$ but the third moment is random. 

Lastly, for $m=4$, I construct a distribution where the first three moments are deterministic. An easy way to keep the first and third moments constant is to make the distribution symmetric such that $\xi_{x,t}(i) = \xi_{x,t}(-i)$. In this spirit, I define the \emph{random three step} distribution as follows. I construct this distribution to have a constant diffusion coefficient of $D=5$. To sample this distribution, I draw a random integer, $X$, uniformly on the interval  $[4, k]$ for $k \in \Z_{>4}$. Note that the distribution is determined by the choice of $k$. I then set $\xi_{x,t}(X) = \xi_{x,t}(-X) = 5 / X^2$ and $\xi_{x,t}(0) = 1 - 10 / X^2$. Note that I do not allow $X$ to be less than four, because if $X < 4$, then $\xi_{x,t}$ cannot be both a probability distribution and have a diffusion coefficient of $D=5$. Since the $\xi_{x,t}$ is always symmetric, the first and third moments are zero. One can easily check that $D=5$ regardless of the choice of $X$. However, the fourth moment is random. I study the random three step distribution for $k=10$ and $k=15$. 

%
%
%
%
\clearpage
\onecolumngrid
\setcounter{equation}{0}
\renewcommand{\theequation}{S\arabic{equation}}
{\centering {\huge \textbf{Supplemental Material} \par}}

\setcounter{section}{0}

\section{Overview of Derivation}

Here, I provide an outline of my strategy to show convergence of the first two moments of the tail probability to the first two moments of the multiplicative stochastic heat equation. This derivation generalizes the methods in \cite{hassUniversalKPZFluctuations2025} but follows in parallel. Thus, I defer to \cite{hassUniversalKPZFluctuations2025} for a more pedagogical approach. 

Before continuing, it is important to more rigorously define some properties of RWRE models. For convenience, I will introduce the notation for the jump distribution in the average environment, $\meanOmega{i} = \expAnnealed{\xi_{x,t}(i)}$. I now define the \emph{annealed} probability measure of $N$ independent random walks, $R^1, \ldots, R^N$ as $\mathbf{P}(\bullet) = \expAnnealed{\mathbb{P}^{\env}(\bullet)}$. This probability measure is simply the probability measure $\mathbb{P}^{\env}$ averaged over all environments, $\env$, sampled via $\randMeasure$. I define the expectation value and variance with respect to the annealed measure as $\mathbf{E}[\bullet]$ and $\mathbf{Var}(\bullet)$. As discussed in the main text, this mean and variance corresponds to averaging a random variable over many samples from many different environments. I define the $k$-\emph{point motion} as $R^1, \ldots, R^k$ for $k \leq N$ with the measure $\mathbf{P}$. The one-point motion is a single random walk, $R^1$, where $R^1(t+1) = R^1(t) + i$ with probability $\meanOmega{i}$. The two-point motion is two random walks, $(R^1, R^2)$, which are i.i.d. when they are apart but are coupled when they are at the same site. When $R^1(t) \neq R^2(t)$, $R^1(t+1) = R^1(t) + i$ with probability $\meanOmega{i}$ and $R^2(t+1) = R^2(t) + j$ with probability $\meanOmega{j}$. In other words, when $R^1(t) \neq R^2(t)$, the random walks behave as if they were in the average environment. When $R^1(t) = R^2(t) = x$, then $R^1(t+1) = x+i$ and $R^2(t+1) = x + j$ with probability $\expAnnealed{\xi_{x,t}(i) \xi_{x,t}(j)}$. 

Rather than studying the tail probability directly, it is easier to work with a more general setup in which the tail probability is a special case. This general setup is the probability mass distribution smoothed against a spatial test function. I include this smoothing  because the probability mass distribution is too noisy to work with directly. The probability mass distribution decays like a Gaussian for large times, so I rescale to ensure the fluctuations remain finite. Putting these elements together, I study 
\begin{equation}\label{eq:UDef}
\mathscr{U}_N(T, \phi) \coloneqq \sum_{x \in \Z} C\left(N, T, \frac{x - \centeringTerm T}{\sqrt{2DN}} \right)  \phi\left(\frac{x - \centeringTerm T}{\sqrt{2DN}}\right) P^{\env}(x, NT )
\end{equation}
where $\phi : \mathbb{R}\rightarrow \mathbb{R}$ is a spacial test function,  
\begin{equation}\label{eq:CDef}
C(N, T, X) \coloneqq \frac{\mathrm{exp}\left\{\displaystyle\frac{\centeringTerm}{2D N^{1/4m}} T + \frac{1}{\sqrt{2D}} N ^{\frac{2m -1}{4m}} X\right\}}{\left(\displaystyle\sum_{i\in\Z} \meanOmega{i} \displaystyle \mathrm{exp}\left\{ \frac{i}{2DN^{1/4m}}\right\} \right)^{NT}} 
\end{equation}
encodes the long time Gaussian behavior where $m-1$ is the number of deterministic moments of the environment and 
\begin{equation*}
\centeringTerm \coloneqq N \frac{ \displaystyle \sum_{j\in\Z} j \meanOmega{j} \mathrm{exp}\left\{ \frac{j}{2D N^{\frac{1}{4m}}} \right\}}{\displaystyle \sum_{i \in \Z} \meanOmega{i} \mathrm{exp}\left\{ \frac{i}{2D N^{\frac{1}{4m}}} \right\}}.
\end{equation*}
By centering the spatial test function at $\centeringTerm T$ and dividing by $\sqrt{2DN}$, I probe the probability distribution at $\centeringTerm T$ around a window of size $\sqrt{2DN}$. It's useful to express Eq. \ref{eq:UDef} as an expectation over an environment $\env$, such that
\begin{equation}\label{eq:UDefExp}
 \mathscr{U}_N(T, \phi) \coloneqq \mathbb{E}^{\env}\left[ C\left(N, T, \frac{R^1(NT) - \centeringTerm T}{\sqrt{2DN}} \right)  \phi\left(\frac{R^1(NT) - \centeringTerm T}{\sqrt{2DN}}\right) \right].
\end{equation}
where $R^1(NT)$ is a random walk in the environment $\env$. 

I will show $\expAnnealed{\mathscr{U}_N(T, \phi)}$ and $\expAnnealed{\mathscr{U}_N(T, \phi)^2}$ converge to the first and second moment of the SHE. The $k^\mathrm{th}$ moment of the SHE is given in terms of the expectation of the local time of $k$ independent Brownian motions \cite{kardarReplicaBetheAnsatz1987, bertiniStochasticHeatEquation1995a} 
\begin{align} \label{eq:SHEmoments}
\mathbb{E}\left[\left(\int_{\mathbb{R}}\phi(X)Z(X,T)dX\right)^k\right] = \mathbf{E}\left[\Phi(\vec{B})\exp\left\{D_0 \sum_{i<j} L^{B^i-B^j}(T) \right\}\right]
\end{align}
where $\mathbb{E}$ on the left is the expectation over the noise $\eta$ of the SHE, and on the right $\Phi(\vec{x}) := \phi(x_1)\cdots\phi(x_k)$, and the expectation $\mathbf{E}$ is over independent Brownian motions $B^1,\ldots B^k$ starting at $0$ and $L^{B^i - B^j}(t)$ is their pair local time at zero, defined as follows. The local time of a space-time continuous Brownian motion $B(t)$ with variance $\sigma^2 t$ is 
\begin{equation*}
    L^B(t) \coloneqq \lim_{\epsilon\rightarrow 0^+} \frac{\sigma^2}{2 \epsilon} \int_0^t \mathbbm{1}_{\left\{-\epsilon < B(s) < \epsilon \right\}} ds. 
\end{equation*}

My general strategy will be to absorb the prefactor, $C(N, T, X)$, into the expectation in Eq. \ref{eq:UDefExp} by interpreting it as a tilting of the increments of the random walk. The first moment can be readily identified as that of the SHE after this tilting. The second moment is more nuanced because it depends on two random walks in the same environment. When these random walks are apart, the tilting works, but when they are at the same site their movement is coupled. This coupling introduces a term which is proportional to a discrete version of the local time. I then take a Taylor expansion of the prefactor on the discrete local time. This Taylor expansion deviates from that in \cite{hassUniversalKPZFluctuations2025} because it depends on the number of fluctuating moments of the environment. I conclude by using results from \cite{hassUniversalKPZFluctuations2025} to relate the discrete local time to the local time of continuous Brownian motion. This culminates in identifying the noise strength of the SHE as 
\begin{equation}\label{eq:noiseStrength}
D_0 = \frac{\extCoef}{(m!)^2 (2 D)^{\frac{4m-1}{2}}}
\end{equation} 
where
\begin{equation}\label{eq:lambdaDef}
    \extCoef = \frac{\momentVar}{\sum_{l=0}^{\infty} \absMeasure{l}\mathbf{E}_\nu[\diffRandomWalk{t+1} - \diffRandomWalk{t} \mid \diffRandomWalk{t} = l]}
\end{equation}
where
\begin{align}\label{eq:absMeasureDef}
\begin{split}
    \momentVar &= \expAnnealed{\left( \sum_{i \in \Z} \xi_{t,x}(i) i^m \right)^2} - \left( \sum_{i \in \Z} \meanOmega{i} i^m \right)^2, \\ 
    \absMeasure{l} &=  \begin{cases} 
        1 & \text{if $l=0$} \\
       2 \mu(l) &  \text{if $l >0$} 
    \end{cases}
\end{split}
\end{align}
and $\invMeasure{l}$ is the unique invariant measure, up to a constant, of the difference random walk $R^1(t) - R^2(t)$, $\diffRandomWalk{t} = |R^1(t) - R^2(t)|$ and $Y \sim \xi_{t,x}$ is a random variable that can be thought of as a single step of a random walk $R^1$. Although the form of $\extCoef$ is rather complex, it simplifies significantly for a number of distributions, which I discuss in the main text and in more detail in \cite{hassExtremeDiffusionMeasures2024a}.

After showing the convergence of the first and second moments of $\mathscr{U}_N(T, \phi)$, I convert these results to show the tail probability also converges to the SHE except with a modified prefactor. 

\section{Convergence of the first moment}
I show the first moment of $\mathscr{U}(T, \phi)$ converges, as $N\rightarrow\infty$ to the first moment of the SHE. I begin by taking the average over all environments of Eq. \ref{eq:UDefExp},
\begin{equation*}
 \expAnnealed{\mathscr{U}_N(T, \phi)} = \mathbf{E}\left[ C\left(N, T, \frac{R^1(NT) - \centeringTerm T}{\sqrt{2DN}} \right)  \phi\left(\frac{R^1(NT) - \centeringTerm T}{\sqrt{2DN}}\right) \right].
\end{equation*}
where $R^1$ is now the one-point motion. I now interpret the coefficient as a tilting of the increments of the random walk $R^1(NT)$. Simplifying the coefficient, I find
\begin{equation*}
C\left(N, T, \frac{R(NT) - \centeringTerm T}{\sqrt{2DN}}\right) = \frac{\exp\left\{ \displaystyle\frac{R(NT)}{2DN^{1/4m}} \right\}}{\left(\displaystyle\sum_{i\in\Z} \meanOmega{i} \displaystyle \mathrm{exp}\left\{ \frac{i}{2DN^{1/4m}}\right\} \right)^{NT}} = \frac{\displaystyle \prod_{i=0}^{NT-1} \exp\left\{\frac{Y_i}{2 D N^{1/4m}}\right\} }{\left(\displaystyle\sum_{i\in\Z} \meanOmega{i} \displaystyle \mathrm{exp}\left\{ \frac{i}{2DN^{1/4m}}\right\} \right)^{NT}}
\end{equation*}
after breaking up the random walk into its increments, $R(NT) = \sum_{i=0}^{NT-1} Y_i$, where $Y_i$ are independent and identically distributed according to $\bar{\xi}$. Notice that there are $NT$ terms in the numerator and denominator. Thus, I can interpret each term as a tilting of the increments of the random walk. The increments of the tilted random walk are independent and identically distributed where now $Y_i = j$ for $j\in\Z$ with probability
\begin{equation*}
\frac{\meanOmega{i} \mathrm{exp}\left\{ \displaystyle \frac{i}{2D N^{1/4m}}\right\}}{\displaystyle \sum_{j\in\Z} \meanOmega{j} \mathrm{exp}\left\{\displaystyle \frac{j}{2D N^{1/4m}} \right\}}.
\end{equation*}
I chose the denominator of $C(N,T,X)$ in Eq. \ref{eq:CDef} so this tilting argument yields increments distributed according to a probability mass distribution.  

Tilting the expectation value to the new probability mass distribution, I find 
\begin{equation}\label{eq:UtiltedExp}
    \expAnnealed{\mathscr{U}_N(T, \phi)} = \tiltedExpAnnealed{\phi\left( \frac{R^1( NT ) -  \centeringTerm T }{\sqrt{2DN}}\right)}.
\end{equation}
where $\tiltedExpAnnealed{\bullet}$ is the tilted expectation value. Under the tilted probability mass distribution, $(R(NT) - \centeringTerm T) / \sqrt{2DN}$ has mean $0$ and variance $T + \mathcal{O}(N^{-1/4m})$ in the limit $N \rightarrow \infty$ where I have dropped higher order terms in the variance. Thus, under the tilted probability mass distribution, the centered and rescaled random walk converges to Brownian motion, i.e.
\begin{equation*}
\lim_{N \rightarrow \infty} \frac{R(NT) - \centeringTerm T}{\sqrt{2DN}} \rightarrow B(T)
\end{equation*}
where $B(T)$ is standard Brownian motion at time $T$. Substituting this into Eq. \ref{eq:UtiltedExp}, I find 
\begin{equation*}
    \lim_{N \rightarrow \infty} \expAnnealed{\mathscr{U}_N(T, \phi)} = \mathbf{E}\left[{\phi\left(B(T)\right)}\right]
\end{equation*}
where $\mathbf{E}[\bullet]$ is the expectation value with respect to a Brownian motion starting at position $0$. This is indeed the first moment of the SHE in Eq. \ref{eq:SHEmoments}.

\section{Convergence of the Second Moment}
I now show the second moment of $\mathscr{U}_N(T, \phi)$ converges, as $N\rightarrow \infty$ to the second moment of the SHE. In doing so, I identify the noise strength of the SHE in Eq. \ref{eq:noiseStrength}.

I begin by squaring Eq. \ref{eq:UDefExp} and averaging over all environments such that
\begin{align}\label{eq:beforeMSub}
\begin{split}
\expAnnealed{\mathscr{U}_N(T, \phi)^2} &= 
\\ \mathbf{E} & \left[ C\left(N, T, \frac{R^1( NT ) -  \centeringTerm T}{\sqrt{2DN}}\right) C\left(N, T, \frac{R^2( NT ) -  \centeringTerm T}{\sqrt{2DN}}\right) \phi\left(\frac{R^1( NT ) -  \centeringTerm T }{ \sqrt{2DN}} \right) \phi\left(\frac{R^2( NT ) -  \centeringTerm T }{ \sqrt{2DN}} \right)\right].
\end{split}
\end{align}
where $(R^1, R^2)$ is the two point motion. I now simplify the prefactors to find
\begin{equation}\label{eq:secondMomentPrefactor}
C\left(N, T, \frac{R^1( NT ) -  \centeringTerm T}{\sqrt{2DN}}\right) C\left(N, T, \frac{R^2( NT ) -  \centeringTerm T}{\sqrt{2DN}}\right) = \frac{\displaystyle \mathrm{exp}\left\{\frac{R^1( NT ) + R^2( NT )}{2D N^{1/4m}}\right\}}{\displaystyle \left(\sum_{i \in \Z} \meanOmega{i} \mathrm{exp}\left\{\frac{i}{2D N^{1/4m}} \right\}\right)^{2NT}}.
\end{equation}
As in the first moment, I want to interpret this prefactor as a tilting of the increments of the two-point motion. This argument follows if the two random walks are at different sites, but doesn't account for the coupled motion when the walks are together. To see this, I break up the denominator of Eq. \ref{eq:secondMomentPrefactor} into $NT$ terms of 
\begin{equation}\label{eq:shiftedNormalization}
\left(\sum_{i \in \Z} \meanOmega{i} \mathrm{exp}\left\{\frac{i}{2D N^{1/4m}} \right\}\right)^2 = \sum_{i, j \in \Z} \meanOmega{i} \meanOmega{j} \mathrm{exp}\left\{\frac{i + j}{2D N^{1/4m}} \right\}.
\end{equation} 
When $R^1(t) \neq R^2(t)$, this correctly normalizes the terms in the numerator of Eq. \ref{eq:secondMomentPrefactor} so the increments of the tilted random walks obey a probability mass distribution. When $R^1(t) = R^2(t)$, the motion of the random walks is coupled and the increments of the tilted random walks no longer obey a probability mass distribution. This coupling can be accounted for by changing $\meanOmega{i}\meanOmega{j}$ to $\expAnnealed{\xi_{x,t}(i) \xi_{x,t}(j)}$ in Eq. \ref{eq:shiftedNormalization}. This can be accomplished by including an additional term to the prefactors, 
\begin{equation*}
C\left(N, T, \frac{R^1( NT ) -  \centeringTerm T}{\sqrt{2DN}}\right) C\left(N, T, \frac{R^2( NT ) -  \centeringTerm T}{\sqrt{2DN}}\right) \exp\left\{ - g\left(\frac{1}{2 D N^{1/4m}}\right) \sum_{i=0}^{t-1} \mathbbm{1}_{\{R^1(i) = R^2(i) \}} \right\}
\end{equation*}  
where $\mathbbm{1}_{\{x=y\}}$ is the indicator function and 
\begin{equation*}
g(\lambda) = \log\left( \sum_{i, j\in\Z} \expAnnealed{\xi_{t,x}(i) \xi_{t,x}(j)} e^{\lambda(i+j)} \right) - 2 \log\left( \sum_{i \in \Z} \bar{\xi}(i) e^{\lambda i}\right).
\end{equation*} 
The extra term proportional $g(\lambda)$ ensures that when $R^1(t) = R^2(t)$ the increments of the tilted random walks obey a probability mass distribution. Including this factor, I can tilt the expectation value in Eq. \ref{eq:beforeMSub} to the new tilted probability mass distribution defined as follows. When $R^1(t) \neq R^2(t)$, $R^1(t+1) = R^1(t) + i$ and $R^2(t+1) = R^2(t) + j$ with probability 
\begin{equation*}
\frac{\displaystyle \meanOmega{i} \meanOmega{j} \exp\left\{\frac{i+j}{2 D N^{1/4m}}\right\}}{\displaystyle \sum_{k,l\in\Z} \meanOmega{k} \meanOmega{l} \exp\left\{\frac{k + l}{2 D N^{1/4m}}\right\}}.
\end{equation*}
When $R^1(t) = R^2(t) = x$, $R^1(t+1) = x +i$ and $R^2(t+1) = x+j$ with probability 
\begin{equation*}
\frac{\displaystyle \expAnnealed{\xi_{t,x}(i) \xi_{t,x}(j)} \exp\left\{\frac{i+j}{2 D N^{1/4m}}\right\}}{\displaystyle \sum_{k,l\in\Z} \expAnnealed{\xi_{t,x}(k) \xi_{t,x}(l)} \exp\left\{\frac{k + l}{2 D N^{1/4m}}\right\}}.
\end{equation*}
I denote $\tiltedExpAnnealedTwo{\bullet}$ as the expectation with respect to the tilted probability mass distribution. Tilting the expectation value in Eq. \ref{eq:beforeMSub}, I find 
\begin{equation}\label{eq:secondMomentBeforeGExpansion}
\expAnnealed{\mathscr{U}_N(T, \phi)^2} = \tiltedExpAnnealedTwo{\exp\left\{g\left((2D)^{-1} N^{-1/4m}\right)\sum_{i=0}^{ NT  -1}\mathbbm{1}_{\{R^1(i) = R^2(i) \}}\right\} \phi\left(\frac{R^1( NT ) -  \centeringTerm T }{ \sqrt{2DN}} \right) \phi\left(\frac{R^2( NT ) -  \centeringTerm T }{ \sqrt{2DN}} \right)}.
\end{equation}
The next step is to take the Taylor expansion of the exponential in the limit that $N\rightarrow \infty$. Up until now, I have followed the analysis done in \cite{hassUniversalKPZFluctuations2025}, but the expansion of $g(\lambda)$ greatly differs. As I will show in Section \ref{sec:gAsymptotics}, as $N\rightarrow \infty$, 
\begin{equation*}
g\left(\frac{1}{2D N^{1/4m}}\right) = \frac{\momentVar}{(m!)^2 (2D)^{2m} \sqrt{N}} + \mathcal{O}\left( \frac{1}{N^{\frac{2m+1}{4m}}}\right).
\end{equation*}
The fact that the leading order term of this expansion scales like $N^{-1/2}$ is quite important. As I will discuss, this scaling ensures the discrete local time converges to the continuous local time. Substituting this expansion into Eq. \ref{eq:secondMomentBeforeGExpansion} above yields
\begin{equation}\label{eq:secondMomentAfterGExpansion}
\expAnnealed{\mathscr{U}_N(T, \phi)^2} \approx \tiltedExpAnnealedTwo{\mathrm{exp}\left\{\frac{\momentVar}{(m!)^2 (2D)^{2m} \sqrt{N}} \sum_{i=0}^{ NT  -1}\mathbbm{1}_{\{R^1(i) = R^2(i) \}}\right\} \phi\left(\frac{R^1( NT ) -  \centeringTerm T }{ \sqrt{2DN}} \right) \phi\left(\frac{R^2( NT ) -  \centeringTerm T }{ \sqrt{2DN}} \right)}.
\end{equation}

To conclude, I only need to show that the centered and rescaled random walks converge to independent Brownian motion and the discrete local time, $\frac{1}{\sqrt{N}}\sum_{i=0}^{ NT  -1}\mathbbm{1}_{\{R^1(i) = R^2(i) \}}$ converges to the pair local time at zero of two independent Brownian motions. 

As in the first moment, $\left(\frac{R^1( NT ) -  c_N T }{ \sqrt{2DN}}, \frac{R^2( NT ) -  c_N T }{ \sqrt{2DN}} \right)$ will converge to independent Brownian motion in the limit that $N\rightarrow \infty$. The two random walks are independent except when $R^1 = R^2$ where their motion is coupled. However, these correlations become negligible in the limit that $N \rightarrow \infty$ unless they are tuned to become stronger as $N \rightarrow \infty$. This is not the case for my models so the centered and rescaled random walks will converge to independent Brownian motion.

In \cite{hassUniversalKPZFluctuations2025}, I show the discrete local time relates to the pair local time via 
\begin{equation*}
\lim_{N \rightarrow \infty} \frac{1}{\sqrt{2DN}} \sum_{i = 0}^{NT - 1} \mathbbm{1}_{\{R^1(i) = R^2(i) \}} = \frac{1}{\sum_{l=0}^{\infty} \absMeasure{l} \mathbf{E}_{\randMeasure}\left[\diffRandomWalk{t+1} - \diffRandomWalk{t} \mid \diffRandomWalk{t} = l \right]} L_0^{B^1 - B^2}(T)
\end{equation*}
where $\diffRandomWalk{t} = |R^1(t) - R^2(t)|$ and 
\begin{align*}
\absMeasure{l} &=  \begin{cases} 
        1 & \text{if $l=0$} \\
       2 \mu(l) &  \text{if $l >0$} 
    \end{cases}
\end{align*}
where $\mu(l)$ is the invariant measure of the difference random walk $R^1(t) - R^2(t)$. Substituting this into Eq. \ref{eq:secondMomentAfterGExpansion} yields 
\begin{equation*}
\lim_{N\to \infty} \expAnnealed{\mathscr{U}_N(T, \phi)^2} = \tiltedExpAnnealedTwo{\mathrm{exp}\left\{\frac{\extCoef}{(m!)^2 (2 D)^{\frac{4m-1}{2}}} L_0^{B^1 - B^2}(T)\right\} \phi(B^1(T)) \phi(B^2(t))},
\end{equation*}
where $\mathbf{E}$ denotes the expectation value of two independent Brownian motion, $B^1$ and $B^2$, starting at position 0 and $\extCoef$ is defined in Eq. \ref{eq:lambdaDef}. This is indeed the second moment of the SHE in Eq. \ref{eq:SHEmoments} with noise strength given by Eq. \ref{eq:noiseStrength}.

\subsection{Expansion of $g(\lambda)$}\label{sec:gAsymptotics}
I now expand $g(\lambda)$ in the limit that $\lambda \rightarrow 0$. My strategy will be to rewrite $g$ in terms of the cumulant and joint cumulant generating function of two random variables. I will then use properties of the cumulant generating function to simplify this expression. Finally, I will evaluate the simplified expression of $g$ for $m=1,2$ and $3$ to observe a pattern and generalize this for all $m$. 

I begin by recalling some useful properties of cumulants and joint cumulants. Let $X$ be a random variable with probability distribution $\mathbb{P}$ and expectation value $\mathbb{E}$. Then the Taylor series of the cumulant generating function, $K(\lambda)$, of $X$ is given by
\begin{equation}\label{eq:cumulantGenerating}
K(\lambda) = \log \left(\mathbb{E}\left[e^{\lambda X}\right] \right) = \sum_{n=1}^{\infty} \frac{\lambda^n}{n!} \kappa_n(X)
\end{equation}
where
\begin{equation*}
\kappa_n(X) = \frac{d^n K(\lambda)}{d\lambda^n} \Bigg|_{\lambda = 0}
\end{equation*}
is the $n^\mathrm{th}$ cumulant of $X$.

I now consider $n$ random variables, $X_1, \ldots, X_n$ with probability measure $\mathbb{P}$ and expectation $\mathbb{E}$. Then the multivariate cumulant generating function, of $X_1, \ldots, X_n$ is given by 
\begin{equation*}
K(\lambda_1, \ldots, \lambda_n) = \log\left( \mathbb{E}\left[ \exp\left\{\sum_{j=1}^N \lambda_i X_i\right\} \right]\right).
\end{equation*}
The joint cumulant of $X_1, \ldots, X_n$ is defined as 
\begin{equation*}
\kappa(X_1, \ldots, X_n) = \frac{d^n}{d\lambda_1 \cdots d\lambda_n} K(\lambda_1, \ldots, \lambda_n).
\end{equation*}
The $n^\mathrm{th}$ cumulant of the addition of two random variables can be expressed in terms of the joint cumulant of copies of the two random variables via 
\begin{equation}\label{eq:addCumulant}
\kappa_n(X_1+X_2) = \sum_{j=0}^n \binom{n}{j} \kappa(\underbrace{X_1, \ldots, X_1}_n, \underbrace{X_2, \ldots, X_2}_{n-j}).
\end{equation}

It is convenient to first express $g(\lambda)$ in terms of the increments of the two-point motion. Let $Y_1, Y_2 \in \Z$ be the increments of the two point motion, $(R^1, R^2)$, when $R^1=R^2$ under the measure $\mathbf{P}$. Then $g(\lambda)$ can be rewritten as 
\begin{equation}\label{eq:randomValueGDef}
g(\lambda) = \log\left( \mathbf{E}\left[ e^{\lambda(Y_1+Y_2)}\right] \right) - 2 \log\left(  \mathbf{E}\left[e^{\lambda Y_1}\right]\right).
\end{equation} 
In the first term on the right hand side, $Y_1$ and $Y_2$ are under the measure of the two-point motion such that $Y_1 = i$ and $Y_2 = j$ with probability $\expAnnealed{\xi_{x,t}(i) \xi_{x,t}(j)}$. Only $Y_1$ appears in the second term, so $Y_1$ in that term is under the measure of the one-point motion such that $Y_1 = i$ with probability $\meanOmega{i}$. Notice that $\log\left(\mathbf{E}\left[e^{\lambda Y_1}\right] \right)$ is the cumulant generating function of $Y_1$ and $\log\left(\mathbf{E}\left[e^{\lambda(Y_1 + Y_2)}\right] \right)$is the joint cumulant generating function of $Y_1$ and $Y_2$.

Substituting the definition of the cumulant generation function in Eq. \ref{eq:cumulantGenerating} into $g(\lambda)$ in Eq. \ref{eq:randomValueGDef}, I find 
\begin{equation}\label{eq:gLambdaJoint}
g(\lambda) = \sum_{n=1}^\infty \frac{\lambda^n}{n!} \kappa_n(Y_1 + Y_2) - 2 \sum_{n=1}^\infty \frac{\lambda^n}{n!} \kappa_n(Y_1).
\end{equation}
I now use Eq. \ref{eq:addCumulant} to see 
\begin{align*}
g(\lambda) &= \sum_{n=1}^\infty \frac{\lambda^n}{n!} \sum_{j=0}^n \binom{n}{j} \kappa(Y_1,\ldots ,Y_2) - 2 \sum_{n=1}^{\infty} \frac{\lambda^n}{n!} \kappa_n(Y_1) \\
&= \sum_{n=1}^{\infty} \frac{\lambda^n}{n!} \left[\sum_{j=0}^n \binom{n}{j} \kappa(Y_1, \ldots, Y_2) - 2 \kappa_n(Y_1) \right]
\end{align*}  
after combining the sums over $n$. The $2 \kappa_n(Y_1)$ term can be canceled by writing out the $j=0$ and $j=n$ terms of the sum over $j$. Doing so I find, 
\begin{equation*}
g(\lambda) = \sum_{n=1}^{\infty} \frac{\lambda^n}{n!} \left[\kappa(Y_1, \ldots ,Y_1) + \kappa(Y_2, \ldots, Y_2) + \sum_{j=1}^{n-1} \binom{n}{j} \kappa(Y_1, \ldots, Y_2) - 2 \kappa_n(Y_1) \right]
\end{equation*}
Notice $\kappa(Y_1, \ldots ,Y_1) = \kappa(Y_2, \ldots, Y_2)$ because $Y_1$ and $Y_2$ are no longer in the same expectation value so they are identically distributed according to $\bar{\xi}$. Furthermore, the $n^\mathrm{th}$ cumulant is equal to the joint cumulant of $n$ copies of $Y_1$, i.e. $\kappa_n(Y_1) = \kappa(Y_1, \ldots, Y_1)$.
Using these properties, I find 
\begin{equation}\label{eq:simplifiedGLambda}
g(\lambda) = \sum_{n=2}^\infty \frac{\lambda^n}{n!} \sum_{j=1}^{n-1}\binom{n}{j} \kappa(\underbrace{Y_1,\ldots, Y_1}_j, \underbrace{Y_2, \ldots, Y_2}_{n-j}).
\end{equation}
where the sum over $j$ now runs over $j=1$ to $j=n-1$ and I have dropped the $n=1$ term because it is equal to $0$. Further simplification of $g(\lambda)$ proves difficult so I calculate Eq. \ref{eq:simplifiedGLambda} for various $m$. 

For $m=1$, I find 
\begin{equation*}
	\lim_{\lambda \rightarrow 0} g(\lambda) = \lambda^2 \left(\mathbf{E}[Y_1 Y_2] - \mathbf{E}[Y_1] \mathbf{E}[Y_2] \right) + \mathcal{O}\left( \lambda^3 \right).
\end{equation*}
It's now useful to write out the expectation values, which yields
\begin{equation*}
\lim_{\lambda \rightarrow 0} g(\lambda) = \lambda^2 \left(\sum_{i, j \in \Z} \expAnnealed{\xi_{x,t}(i)\xi_{x,t}(j)} i j - \sum_{i, j \in \Z} \meanOmega{i} \meanOmega{j} i j \right) + \mathcal{O}\left(\lambda^3\right).
\end{equation*}
Note, the second term is $0$ since I only consider net drift free systems. Now let $Y \sim \xi_{x,t}$ such that, 
\begin{equation*}
\lim_{\lambda \rightarrow 0} g(\lambda) = \lambda^2 \varAnnealed(\expEnv[Y]) + \mathcal{O}\left(\lambda^3\right)
\end{equation*}
where $\varAnnealed(\expEnv[Y]) = \sum_{i, j \in \Z} \expAnnealed{\xi_{x,t}(i)\xi_{x,t}(j)} i j$. This agrees with the expansion of $g(\lambda)$ in \cite{hassUniversalKPZFluctuations2025} which studied the $m=1$ case. 

I now consider the case when $m=2$ where the first moment $\sum_{i \in \Z} \xi_{t,x}(i) i = 0$ with probability 1. Using this property, I find 
\begin{equation*}
\mathbf{E}[Y_1 Y_2^n] = \sum_{i, j \in \Z} \expAnnealed{\xi_{x,t}(i) \xi_{x,t}(j)} i j^n = \expAnnealed{\left(\sum_{i\in\Z} \xi_{x,t}(i) i\right) \left(\sum_{j\in\Z} \xi_{x,t}(j) j^n \right)}= 0
\end{equation*}
for all $n \in \Z_{\geq 1}$ after separating the sums. With these terms equal to $0$, I find  
\begin{align*}
\lim_{\lambda \rightarrow 0} g(\lambda) &= \frac{\lambda^4}{4!} \binom{4}{2}\left( \mathbf{E}_\randMeasure[Y_1^2 Y_2^2] - \mathbf{E}_\randMeasure[Y_1^2]^2 \right) + \mathcal{O}\left( \lambda^5 \right)
\\ 
&= \frac{\lambda^4}{4!} \binom{4}{2}\left( \sum_{i, j \in \Z} \expAnnealed{\xi_{x,t} (i) \xi_{x,t}(j)} i^2 j^2 - \sum_{i, j\in\Z} \meanOmega{i} \meanOmega{j} i^2 j^2 \right) + \mathcal{O}\left( \lambda^5 \right)
\end{align*}
As in the first moment this expression can be made more compact by writing it in terms of $Y\sim\xi_{x,t}$. Doing so yields
\begin{equation*}
\lim_{\lambda \rightarrow 0} g(\lambda) = \frac{\lambda^4}{(2!)^2}\mathrm{Var}_\randMeasure(\expEnv[Y^2]) + \mathcal{O}\left( \lambda^5 \right)
\end{equation*}
where I've used $\binom{4}{2} = 4! / (2!)^2$ and $\mathrm{Var}_\randMeasure(\expEnv[Y^2]) = \sum_{i, j \in \Z} \expAnnealed{\xi_{x,t} (i) \xi_{x,t}(j)} i^2 j^2 - \sum_{i, j\in\Z} \meanOmega{i} \meanOmega{j} i^2 j^2$. 

I now consider the $m=3$ where $\mathbf{E}[Y_1 Y_2^n] = 0$ for all $n \in \Z_{\geq 1}$ as well as a constant second moment such that $\sum_{i\in\Z} \xi_{x,t}(i) i^2 = 2D$ with probability $1$ where $D$ is the diffusion coefficient in Eq. \ref{eq:diffusionCoefficient}. Thus,
\begin{equation*}
\mathbf{E}[Y_2^2 Y_2^n] = \expAnnealed{\left(\sum_{i\in\Z} \xi_{x,t}(i) i^2\right) \sum_{j \in \Z} \xi_{x,t}(j)j^n} = 2 D \mathbf{E}[Y_2^n]
\end{equation*} 
for all $n \in \Z_{\geq 1}$. As expected, $g(\lambda)$ simplifies to
\begin{equation*}
\lim_{\lambda \rightarrow 0} g(\lambda) = \frac{\lambda^6}{(3!)^2} \mathrm{Var}_\randMeasure(\expEnv[Y^3]) + \mathcal{O}\left( \lambda^7 \right)
\end{equation*}
Thus, I extrapolate to higher $m$,
\begin{equation}\label{eq:asymptoticGLambda}
\lim_{\lambda \rightarrow 0} g(\lambda) =  \frac{\lambda^{2m}}{(m!)^2} \varAnnealed\left(\expEnv[Y^{m}] \right) + \mathcal{O}\left( \lambda^{2m+1} \right)
\end{equation}
as desired.

\section{Convergence of the Tail Probability} 
To derive the distribution of the extreme location and extreme first passage time, I convert the probability mass distribution, to the tail probability 
\begin{equation}\label{eq:tailProbabilityDef}
\mathbb{P}^{\env}\left(R(NT) \geq \centeringTerm T + \sqrt{2DN} X\right) = \sum_{y \geq \centeringTerm T + \sqrt{2DN} X} \mathbb{P}^{\env}\left(R(NT) = y\right).
\end{equation}
In \cite{hassUniversalKPZFluctuations2025}, I do this for the $m=1$ case and show the tail probability is also characterized by the SHE but with a modified prefactor. The derivation for higher $m$ is nearly identical so I will only provide an overview of the derivation here. 

The tail probability in Eq. \ref{eq:tailProbabilityDef} is a special case of $\mathscr{U}_N(T, \phi)$ where 
\begin{equation*}
\phi_{\mathrm{tail}}\left(X'\right) = \frac{1}{C(N,T,X)}\exp\left\{ -\frac{N^{\frac{2m-1}{4m}}}{\sqrt{2D}}(X'-X) \right\} \mathbbm{1}_{\left\{X' \geq X\right\}}
\end{equation*}
such that 
\begin{equation}\label{eq:tailToProbDist}
\mathbb{P}^{\env}\left(R(NT) \geq \centeringTerm T + \sqrt{2DN} X\right) = \mathscr{U}_N(T, \phi_{tail}).
\end{equation}
Assuming convergence of $\mathscr{U}_N(T, \phi)$ to the SHE as $N \rightarrow \infty$, yields
\begin{align*}
\mathbb{P}^{\env}\left(R(NT) \geq \centeringTerm T + \sqrt{2DN} X\right) &= \int_{\mathbb{R}} \phi_{\mathrm{tail}}(X') Z(X', T) dX' \\
&= \frac{1}{C(N,T,X)} \int_{\mathbb{R}} \exp\left\{ -\frac{N^{\frac{2m-1}{4m}}}{\sqrt{2D}}(X'-X) \right\} \mathbbm{1}_{\left\{X' \geq X\right\}} Z(X', T) dX'
\end{align*}
where I've substituted the definition of $\phi_{\mathrm{tail}}$ in the second line. Notice that for $X' > X$, the exponential in the integral vanishes as $ N \rightarrow \infty$. Due to the indicator function, I only integrate over the random where $X' \geq X$. Thus, I approximate the exponential in the limit that $N \rightarrow \infty$ as  
\begin{equation*}
\exp\left\{ -\frac{N^{\frac{2m-1}{4m}}}{\sqrt{2D}}(X'-X) \right\} \approx \delta \left( -\frac{N^{\frac{2m-1}{4m}}}{\sqrt{2D}}(X'-X) \right)
\end{equation*}
where $\delta(x)$ is the Dirac delta function. Making this approximation, I evaluate the integral above to find 
\begin{equation*}
\mathbb{P}^{\env}\left(R(NT) \geq \centeringTerm T + \sqrt{2DN} X\right) = \frac{\sqrt{2D}}{N^{\frac{2m-1}{4m}}C(N, T, X)} Z(X, T)
\end{equation*}
as $N \rightarrow \infty$. After bringing the prefactor to the left hand side,
\begin{equation}\label{eq:tailProbability}
\lim_{N \rightarrow \infty} \frac{N^{\frac{2m-1}{4m}}C(N, T, X)}{\sqrt{2D}} \mathbb{P}^{\env}\left(R(NT) \geq \centeringTerm T + \sqrt{2DN} X\right) =  Z(X, T).
\end{equation} 
Therefore, the tail probability converges to the SHE with the same noise strength in Eq. \ref{eq:noiseStrength}, but with a different prefactor.  

\section{Asymptotic Expansion of the Tail Probability}
Before beginning the analysis of the extreme location and first passage time, I first change variables in the tail probability to consider a random walk at time $t$. In contrast to a random walk at time $NT$ in Eq. \ref{eq:tailProbability}. 

I begin by taking the $\ln$ of both sides of Eq. \ref{eq:tailProbability} and solving for tail probability, such that 
\begin{align}\label{eq:tailProbApprox}
\begin{split}
	\ln\left(\mathbb{P}^{\env}\left( R(NT) \geq \centeringTerm T  +  \sqrt{2DN} X\right)\right) = & -\frac{\centeringTerm T}{2 D N^{1/4m}} - \frac{N^{\frac{2m-1}{4m}} X}{\sqrt{2D}} + NT \ln \left( \sum_{i \in \Z} \meanOmega{i} \mathrm{exp}\left\{ \frac{i}{2D N^{1/4m}}\right\} \right) \\ 
	& - \ln\left( \frac{N^{\frac{2m-1}{4m}}}{\sqrt{2D}} \right) + h(X, T)
\end{split}
\end{align}
where $h(X, T) = \ln(Z(X, T))$ now solves the KPZ equation with narrow wedge initial condition and noise strength given by Eq. \ref{eq:noiseStrength}. I now substitute $t \coloneqq NT$ and $v \coloneqq T^{1/4m}$ to find 
\begin{align}\label{eq:tailWithVT}
\begin{split}
	\ln\left(\mathbb{P}^{\env}\left( R(t) \geq c\left(\frac{t}{v^{4m}}\right) v^{4m} + \sqrt{\frac{2Dt}{v^{4m}}}  X\right)\right) = & -\frac{c\left(\frac{t}{v^{4m}}\right) v^{4m+1}}{2 D t ^{1/4m}} - \frac{t^{\frac{2m-1}{4m}} X}{v^{2m-1}\sqrt{2D}} + t \ln \left( \sum_{i \in \Z} \meanOmega{i} \mathrm{exp}\left\{ \frac{i v}{2D t^{1/4m}}\right\} \right) \\ 
	& - \ln\left( \frac{t^{\frac{2m-1}{4m}}}{v^{2m-1}\sqrt{2D}} \right) + h(X, v^{4m}).
\end{split}
\end{align} 
To simplify, I expand in the limit that $t \rightarrow \infty$. I will only track the lowest order terms in $t$ as higher order terms do not enter the extreme location and extreme first passage time calculations. Expanding the function $c(t / v^{4m})$, I find 
\begin{align*}
c\left( \frac{t}{v^{4m}}\right) &= \frac{t}{v^{4m}} \frac{ \displaystyle \sum_{j\in\Z} j \meanOmega{j} \mathrm{exp}\left\{ \frac{v j}{2D t^{1/4m}} \right\}}{\displaystyle \sum_{i \in \Z} \meanOmega{i} \mathrm{exp}\left\{ \frac{v i}{2D t^{1/4m}} \right\}} \\
&\approx \frac{t}{v^{4m}} \frac{\displaystyle \sum_{j \in \Z} j \meanOmega{j} \left(1 + \frac{vj}{2Dt^{1/4m}}\right)}{\displaystyle \sum_{i \in \Z} \meanOmega{i} \left(1 + \mathcal{O}\left(t^{-1/4m}\right)\right)} \\
\end{align*}
where I've used the approximation $e^{x} \approx 1 + x$ as $x \rightarrow 0$. Since $\bar{\xi}$ is a probability distribution on $\Z$, $\sum_{i \in \Z} \meanOmega{i} = 1$. Furthermore, since I only consider distributions with a zero mean in the average environment, $\sum_{j \in \Z} \meanOmega{j} j = 0$. Using these properties, I simplify
\begin{align*}
c\left(\frac{t}{v^{4m}} \right) &\approx \frac{t}{v^{4m}} \left( \frac{v}{2D t^{1/4m}} \sum_{j \in \Z}\meanOmega{j} j^2 \right) \\
&\approx \frac{t^{\frac{4m-1}{4m}}}{v^{4m-1}}.
\end{align*}
where I've dropped higher order terms in $t$ and used $\sum_{i \in \Z} \meanOmega{i} i^2 = 2D$. 

I also expand the logarithmic term such that
\begin{align*}
\ln \left( \sum_{i \in \Z} \meanOmega{i} \mathrm{exp}\left\{ \frac{i v}{2D t^{1/4m}}\right\} \right) &\approx \ln \left( \sum_{i \in \Z} \meanOmega{i} \left(1 + \frac{i v}{2 D t^{1/4m}} + \frac{ i^2 v^2}{2(2D)^2 t^{1/2m}} \right) \right) \\
&= \ln\left(\sum_{i \in \Z} \meanOmega{i} + \frac{v}{2D t^{1/4m}}\sum_{i \in \Z} \meanOmega{i} i + \frac{v^2}{2(2D)^2 t^{1/2m}} \sum_{i \in \Z} \meanOmega{i} i^2\right) \\
\end{align*}
where I've used the small argument expansion of the exponential. As before, I use the properties $\sum_{i \in \Z} \meanOmega{i} = 1,\ \sum_{i \in \Z} \meanOmega{i} i = 0$, and $\sum_{i \in \Z} \meanOmega{i} i^2 = 2D$ to simplify
\begin{align*}
\ln \left( \sum_{i \in \Z} \meanOmega{i} \mathrm{exp}\left\{ \frac{i v}{2D t^{1/4m}}\right\} \right) &= \ln\left( 1 + \frac{v^2}{4D t^{1/2m}}\right) \\
&\approx \frac{v^2}{4D t^{1/2m}}.
\end{align*}
where I've used $\ln(1+x) \approx x$ for $x \rightarrow 0$. 

Substituting these approximations into Eq. \ref{eq:tailWithVT}, I find 
\begin{equation*}
\ln\left( \mathbb{P}^{\env} \left(R(t) \geq vt^{\frac{4m-1}{4m}} + \sqrt{\frac{2Dt}{v^{4m}}} X\right)\right) \approx - \frac{v^2 t^{\frac{2m-1}{2m}}}{4D} - \frac{t^{\frac{2m-1}{4m}}X}{v^{2m-1}\sqrt{2D}} + h(X, v^{4m})
\end{equation*}
where I've dropped the term proportional to $\ln(t)$.  
 
Recall $h(X, T)$ has noise strength given by Eq. \ref{eq:noiseStrength}. I now transform the space and time variables of the KPZ equation such that the noise strength is $1$. I define $\tilde{h}(\tilde{X}, \tilde{T}) = h(X, T)$ with $\tilde{X} = \frac{2 \extCoef}{(m!)^2 (2D)^{\frac{4m-1}{2}}} X$ and $\tilde{T} = \frac{4 \extCoef^2}{(m!)^4 (2D)^{4m-1}} T$ where $\tilde{h}$ solves the standard coefficient KPZ equation 
\begin{equation}\label{eq:standardKPZ}
	\partial_{\tilde{T}} \tilde{h} = \frac{1}{2} \partial_{\tilde{X}}^2 \tilde{h} + \frac{1}{2} \left(\partial_{\tilde{X}} \tilde{h}\right)^2 + \eta .
\end{equation}
Making these transformations, I find 
\begin{equation}\label{eq:tailProbFull} 
\ln\left( \mathbb{P}^{\env} \left(R(t) \geq vt^{\frac{4m-1}{4m}} + \sqrt{\frac{2Dt}{v^{4m}} X}\right)\right) \approx - \frac{v^2 t^{\frac{2m-1}{2m}}}{4D} - \frac{t^{\frac{2m-1}{4m}}X}{v^{2m-1}\sqrt{2D}} + \tilde{h}\left(\frac{2 \extCoef}{(m!)^2 (2D)^{\frac{4m-1}{2}}} X, \frac{4 \extCoef^2}{(m!)^4 (2D)^{4m-1}} v^{4m}\right).
\end{equation}
Note that I will drop the tilde on $\tilde{h}$ as I will always use the standard coefficient KPZ equation in Eq. \ref{eq:standardKPZ} for now on.

\section{Extreme Position}
Here I derive the large time asymptotics of the extreme location. The analysis follows almost identically from \cite{hassExtremeDiffusionMeasures2024a, hassAnomalousFluctuationsExtremes2023} so I provide a summary of the methods here, but refer to \cite{hassExtremeDiffusionMeasures2024a, hassAnomalousFluctuationsExtremes2023} for a more in-depth treatment. 

I first derive the mean and variance of $\envMax$. Since my definition of $\envMax$ is in terms of $\mathbb{P}^{\env}(R(t) \geq x)$, I first have to transform Eq. \ref{eq:tailProbFull} to match. I do so by substituting $x = v t^\frac{4m-1}{4m}$ and $X=0$ into Eq. \ref{eq:tailProbFull}, such that 
\begin{equation}\label{eq:tailProbInTermsOfL}
 \ln(\mathbb{P}^{\env}(R(t) \geq x)) = - \frac{x^2}{4 D t} + h\left(0, \frac{4 \extCoef^2}{(m!)^4 (2D)^{4m-1}} \frac{x^{4m}}{t^{4m-1}} \right).
\end{equation} 
I now use the definition of the extreme location to set $\mathbb{P}^{\env}(R(t) \geq x) = 1/N$ such that 
\begin{equation}\label{eq:envImplicitEquation}
\ln(N) \approx \frac{(\envMax)^2}{4 D t} - h\left(0, \frac{4 \extCoef^2}{(m!)^4 (2D)^{4m-1}} \frac{(\envMax)^{4m}}{t^{4m-1}} \right).
\end{equation}
Note that Eq. \ref{eq:envImplicitEquation} is an approximation because $\envMax$ need not occur exactly at the position when $\mathbb{P}^{\env}(R(t) \geq x) = 1/N$ but the maximum position satisfying $\mathbb{P}^{\env}(R(t) \geq x) \geq 1/N$.

I approximate $\envMax$ in the asyptotic limit that $t^{2m-1} \gg \extCoef^2 \ln(N)^{2m}$ and $N \gg 1$ by perturbatively expanding Eq. \ref{eq:envImplicitEquation}. Specifically, I find 
\begin{equation*}
\envMax \approx \sqrt{4 D \ln(N) t} + \sqrt{\frac{Dt}{\ln(N)}} h\left( 0, \frac{8 \extCoef^2 \ln(N)^{2m}}{(m!)^4 (Dt)^{2m-1}} \right).
\end{equation*}
Since only the term proportional to the KPZ equation is random, the mean and variance of $\envMax$ are
\begin{align}
\expAnnealed{\envMax} &\approx \sqrt{4D \ln(N) t} \\ 
\varAnnealed\left(\envMax\right) &= \frac{Dt}{\ln(N)} \var{h\left( 0, \frac{8 \extCoef^2 \ln(N)^{2m}}{(m!)^4 (Dt)^{2m-1}} \right)}. \label{eq:varEnvMaxKPZ}
\end{align}
where I have only kept the leading order term of the mean.  
 
For $t^{2m-1} \gg \extCoef^2 \ln(N)^{2m}$, I derive the asymptotic scaling, in time, of $\varAnnealed\left(\envMax\right)$. In this regime, I use the small time expansion of the KPZ equation 
\begin{equation}\label{eq:KPZShortTime}
h(0,s) \approx -\frac{s}{24} - \ln\left( \sqrt{2\pi s}\right) + \left( \frac{\pi s}{4}\right)^{1/4} G_s
\end{equation}
where $G_s$ converges to a standard Gaussian as $s \rightarrow \infty$ \cite{amirProbabilityDistributionFree2011a, calabreseFreeEnergyDistributionDirected2010a}. Substituting this expansion into Eq. \ref{eq:varEnvMaxKPZ}, I find 
\begin{equation*}
\varAnnealed(\envMax) \approx \frac{\sqrt{2 \pi}}{(m!)^2} \extCoef \ln(N)^{m-1} (Dt)^{\frac{3-2m}{2}}
\end{equation*}
for $t^{2m-1} \gg \ln(N)^{2m}$. For $m=1$, $\varAnnealed\left(\envMax\right) = \extCoef \sqrt{2 \pi D t}$, which matches results in \cite{hassExtremeDiffusionMeasures2024a}. 

I now derive the distribution of $\samMax$ and identify the mean and variance. I begin by deriving the tail probability distribution by combining Eq. \ref{eq:minDist} and the definition of $\samMax$ such that 
\begin{equation*}
\mathbb{P}^{\env}(\samMax \leq x) = (1 - \mathbb{P}^{\env}(R(t) \leq x + \envMax))^N
\end{equation*}
I now use the approximation $(1 + x)^N \approx e^{x N}$ for $x \ll 1$ and $N \rightarrow \infty$ to find 
\begin{equation*}
\ln\left( \mathbb{P}^{\env}(\samMax \leq x) \right) \approx - N \mathbb{P}^{\env}(R(t) \leq x + \envMax)
\end{equation*}
after taking the logarithm of both sides. I now substitute Eq. \ref{eq:tailProbInTermsOfL} to find 
\begin{equation*}
\ln\left( \mathbb{P}^{\env}(\samMax \leq x) \right) \approx - N \exp \left(- \frac{(x + \envMax)^2}{4Dt} \right)
\end{equation*}
where I've only kept the leading order term. I also assume $x \ll \envMax$, such that $(x + \envMax)^2 \approx (\envMax)^2 + 2 x\envMax$. As I will see, $\expAnnealed{\envMax} \gg \expAnnealed{\samMax}$ which justifies this approximation. Using this approximation, I find 
\begin{equation*}
\ln\left( \mathbb{P}^{\env}(\samMax \leq x) \right) \approx - N \exp \left( -\frac{(\envMax)^2}{4Dt} - \frac{x \envMax}{2Dt} \right).
\end{equation*}
I now approximate $\envMax$ with its leading order term, $\sqrt{4 D t \ln(N)}$. In making this approximation, I assume that $\samMax$ and $\envMax$ are independent. Although I do not justify this here, my numerics indicate this is a reasonable assumptions. Substituting this approximation, I find 
\begin{equation*}
\ln\left( \mathbb{P}^{\env}(\samMax \leq x) \right) \approx - \exp \left( - \sqrt{\frac{\ln(N)}{Dt}}x \right) 
\end{equation*}
after simplifying. Therefore, $\samMax$ is Gumbel distributed with location $0$ and scale $\sqrt{\frac{Dt}{\ln(N)}}$. Therefore, the mean and variance of $\samMax$ is 
\begin{align}
\expAnnealed{\samMax} &\approx \sqrt{\frac{Dt}{\ln(N)}} \gamma \\
\varAnnealed(\samMax) &\approx \frac{\pi^2 Dt}{6\ln(N)}
\end{align}
where $\gamma \approx 0.577$ is the Euler-Mascheroni constant. Notice that unlike $\envMax$, $\samMax$ does not depend on $m$. 

With the mean and variance of both $\envMax$ and $\samMax$, I finally derive the mean and variance of $\max$. Rearranging my definition of $\samMax$, I see $\max = \samMax + \envMax$. I find to leading order the mean of $\max$ is given by $\expAnnealed{\max} \approx \expAnnealed{\envMax}$ since $\expAnnealed{\envMax} \gg \expAnnealed{\samMax}$. I find the variance of $\max$ is given by $\varAnnealed(\max) \approx \varAnnealed(\samMax) + \varAnnealed(\envMax)$ since $\samMax$ and $\envMax$ are independent. Using my asymptotic approximations of the mean and variance of $\envMax$ and $\samMax$, I find 
\begin{align}
\expAnnealed{\max} &\approx \sqrt{4Dt \ln(N)} \\
\varAnnealed(\max) &\approx \frac{\pi^2 Dt}{6 \ln(N)} + \frac{\sqrt{2 \pi}}{(m!)^2} \extCoef \ln(N)^{m-1} (Dt)^{\frac{3-2m}{2}}
\end{align}
for $t^{2m-1} \gg \extCoef^2 \ln(N)^{2m}$ and $N \gg 1$.
\section{Extreme First Passage Time}

Here I derive the large distance asymptotics for the extreme first passage time. The analysis follows almost identically from \cite{hassExtremeDiffusionMeasures2024a, hassFirstpassageTimeManyparticle2024} so I provide a summary of the methods here, but refer to \cite{hassExtremeDiffusionMeasures2024a, hassFirstpassageTimeManyparticle2024} for a more in-depth treatment.  

I begin by using the non-backtracking approximation $\mathbb{P}^{\env}(\tau_L \leq t) \approx \mathbb{P}^{\env}(R(t) \geq L)$ to convert the distribution of the first passage time to the distribution of the position of a random walker. This assumes that once a particle passes the location $L$, the particle will not backtrack and cross past position $L$ again. I substitute this approximation into Eq. \ref{eq:tailProbInTermsOfL} and find
\begin{equation}\label{eq:tauLApprox}
\ln(\mathbb{P}^{\env}(\tau_L \leq t)) \approx  - \frac{L^2}{4 D t} + h\left(0, \frac{4 \extCoef^2}{(m!)^4 (2D)^{4m-1}} \frac{L^{4m}}{t^{4m-1}} \right).
\end{equation}
I now use the definition of $\envFPT$ to set $\mathbb{P}^{\env}(\tau_L \leq t) = 1/N$ such that
\begin{equation}\label{eq:fptImplicitEquation}
\ln(N) \approx  \frac{L^2}{4 D \envFPT} - h\left(0, \frac{4 \extCoef^2}{(m!)^4 (2D)^{4m-1}} \frac{L^{4m}}{(\envFPT)^{4m-1}} \right)
\end{equation}
Note that this is an approximation since $\envFPT$ need not occur exactly at the time when $\mathbb{P}^{\env}(\tau_L \leq t) = 1/N$ but the minimum time satisfying $\mathbb{P}^{\env}(\tau_L \leq t) \geq 1/N$.   

I approximate $\envFPT$ in the asymptotic limit that $L^{4m-2} \gg \extCoef^2 \ln(N)^{4m-1}$ and $N \gg 1$ by perturbatively expanding Eq. \ref{eq:fptImplicitEquation}. Specifically, I find 
\begin{equation*}
\envFPT \approx \frac{L^2}{4D\ln(N)} - \frac{L^2}{4D\ln(N)^2} h \left(0, \frac{2^{4m+1}\extCoef^2 \ln(N)^{4m-1}}{(m!)^4 L^{4m-2}} \right).
\end{equation*}
Since only the term proportional to the KPZ equation is random, the mean and variance of $\envFPT$ are 
\begin{align}
\expAnnealed{\envFPT} &\approx \frac{L^2}{4D\ln(N)} \\
\varAnnealed(\envFPT) &\approx \frac{L^4}{(4D)^2 \ln(N)^2} \var{h\left(0, \frac{2^{4m+1}\extCoef^2 \ln(N)^{4m-1}}{(m!)^4 L^{4m-2}} \right)}.\label{eq:envFPTKPZVar}
\end{align}
For $L^{4m-2} \gg \extCoef^2 \ln(N)^{4m-1}$, I derive the asymptotic scaling in position of $\varAnnealed(\envFPT)$ using the small time approximation of the KPZ equation in Eq. \ref{eq:KPZShortTime}. Substituting this expansion into Eq. \ref{eq:envFPTKPZVar}, I find
\begin{equation*}
\varAnnealed(\envFPT) \approx \frac{\extCoef \sqrt{\pi} 2^{\frac{4m-9}{2}}}{(m!)^2 D^2} \ln(N)^{\frac{4m-9}{2}} L^{5 - 2m}.
\end{equation*}
For $m=1$, I see 
\begin{equation*}
\varAnnealed(\envFPT) \approx \extCoef \frac{\sqrt{2\pi} L^3}{8 D^2 \log(N)^{5/2}}
\end{equation*}
which matches results in \cite{hassExtremeDiffusionMeasures2024a}. 

I now derive the distribution of $\samFPT$ and identify the mean and variance. I begin by deriving the tail probability distribution by combining Eq. \ref{eq:minDist} and the definition of $\samFPT$ such that 
\begin{equation*}
\mathbb{P}^{\env}(\samFPT \geq t) = (1 - \mathbb{P}^{\env}(\tau_L \leq t + \envFPT))^N.
\end{equation*}
I now use the approximation $(1 + x)^{N} \approx e^{xN}$ for $x \ll 1$ and $N \rightarrow \infty$ to find 
\begin{equation*}
\ln(\mathbb{P}^{\env}(\samFPT \geq t)) \approx - N \mathbb{P}^{\env}(\tau_L \leq t + \envFPT)
\end{equation*}
after taking the logarithm of both sides. I now substitute Eq. \ref{eq:tauLApprox} which yields
\begin{equation*}
\ln(\mathbb{P}^{\env}(\samFPT \geq t)) \approx - N \exp\left( - \frac{L^2}{4 D (t + \envFPT)} \right)
\end{equation*}
where I've only kept the leading order term. I also assume $t \ll \envFPT$, such that $\frac{1}{t + \envFPT} \approx \frac{1}{\envFPT}(1-\frac{t}{\envFPT}) = \frac{1}{\envFPT} - \frac{t}{(\envFPT)^2}$. As I will show, $\expAnnealed{\envFPT} \gg \expAnnealed{\samFPT}$ which justifies this approximation. Using this approximation, I find 
\begin{equation*}
\ln(\mathbb{P}^{\env}(\samFPT \geq t)) \approx - N \exp\left( - \frac{L^2}{4 D\envFPT} + \frac{t L^2}{4 D(\envFPT)^2} \right). 
\end{equation*}
I now approximate $\envFPT$ with its leading order term $\frac{L^2}{4D\ln(N)}$. In making this approximation, I assume that $\samFPT$ and $\envFPT$ are independent. Although I do not justify this here, my numerics indicate this is a reasonable assumption. Substituting this approximation, I find 
\begin{equation*}
\ln(\mathbb{P}^{\env}(\samFPT \geq t)) \approx - \exp\left( \frac{4D \ln(N)^2}{L^2} t \right) 
\end{equation*}
after simplifying. Therefore, $-\samFPT$ is Gumbel distributed with location $0$ and scale $\frac{L^2}{4 D \ln(N)^2}$. Then, the mean and variance of $\samFPT$ is 
\begin{align*}
\expAnnealed{\samFPT} &\approx - \frac{\gamma L^2}{4D\ln(N)^2} \\
\varAnnealed(\samFPT) &\approx \frac{\pi^2 L^4}{96 D^2 \ln(N)^4}
\end{align*}
where $\gamma$ is the the Euler-Mascheroni constant. Notice that unlike $\envFPT$, $\samFPT$ does not depend on $m$. 

With the mean and variance of both $\envFPT$ and $\samFPT$, I finally derive the mean and variance of $\min$. Rearranging my definition of $\samFPT$, I see $\min = \samFPT + \envFPT$. I find to leading order that the mean of $\min$ is given by $\expAnnealed{\min} \approx \expAnnealed{\envFPT}$ since $\expAnnealed{\envFPT} \gg \expAnnealed{\samFPT}$. I find the variance of $\min$ is given by $\varAnnealed\left(\min\right) \approx \varAnnealed\left(\samFPT\right) + \varAnnealed\left(\envFPT\right)$ since $\samFPT$ and $\envFPT$ are independent. Using my asymptotic approximations of the mean and variance of $\envFPT$ and $\samFPT$, I find 
\begin{align}
\expAnnealed{\min} &\approx \frac{L^2}{4 D \ln(N)} \\
\varAnnealed(\min) & \approx  \frac{\pi^2 L^4}{96 D^2 \ln(N)^4} + \frac{\extCoef \sqrt{\pi} 2^{\frac{4m-9}{2}}}{(m!)^2 D^2} \ln(N)^{\frac{4m-9}{2}} L^{5 - 2m}
\end{align}
for $L^{4m-2} \gg \extCoef^2 \ln(N)^{4m-1}$ and $N \gg 1$.
\end{document}